\documentclass[journal=ancac3,manuscript=article]{achemso}

\pdfoutput=1
\usepackage[version=3]{mhchem} 

\newcommand{\degree}{\ensuremath{^\circ}}

\author{William R. French}
\affiliation[Vanderbilt University]{Department of Chemical and Biomolecular Engineering, Vanderbilt University, Nashville, TN}
\author{Christopher R. Iacovella}
\affiliation[Vanderbilt University]{Department of Chemical and Biomolecular Engineering, Vanderbilt University, Nashville, TN}
\author{Peter T. Cummings}
\email{peter.cummings@vanderbilt.edu}
\affiliation[Vanderbilt University]
{Department of Chemical and Biomolecular Engineering, Vanderbilt University, Nashville, TN}
\alsoaffiliation[Oak Ridge National Laboratory]{Center for Nanophase Materials Sciences, Oak Ridge National Laboratory, Oak Ridge, TN}

\title{Large-Scale Atomistic Simulations of Environmental Effects on the Formation and Properties of Molecular Junctions}

\begin{document}

\pagebreak

\begin{abstract}

Using an updated simulation tool, we examine molecular junctions comprised of benzene-1,4-dithiolate bonded between gold nanotips, focusing on the importance of environmental factors and inter-electrode distance on the formation and structure of bridged molecules.  We investigate the complex relationship between monolayer density and tip separation, finding that the formation of multi-molecule junctions is favored at low monolayer density, while single-molecule junctions are favored at high density.  We demonstrate that tip geometry and monolayer interactions, two factors that are often neglected in simulation, affect the bonding geometry and tilt angle of bridged molecules.  We further show that the structures of bridged molecules at 298 and 77 K are similar.

\end{abstract}

\textsc{}


\vspace{0.5in}


Conductance measurements through molecular junctions have been at the forefront of nanoscale research for over a decade \cite{Tao:2006,Nitzan:2003,Nichols:2010}. This work is motivated by the potential for fabrication of molecular-based electronic circuit elements \cite{Bandyopadhyay:2010} and, perhaps more so, discrepancies in the experimentally \cite{Reed:1997} and theoretically \cite{DiVentra:2000} reported conductance through a single molecule. The discrepancies have improved over the years \cite{Lindsay:2007}, due in part to the development of highly automated and optimized experimental techniques ($e.g.$, scanning tunneling microscopy break junction method \cite{Xu:2003,Venkataraman:2006,Venkataraman-Nature:2006}, nanofabricated mechanically-controllable break junction technique \cite{Tsutsui:2006,Tsutsui:2009}), as well as the emergence of theoretical tools ($e.g.$, self-consistent GW calculations \cite{Strange:2011}, approximate self-interaction corrections \cite{Toher:2007,Pontes:2011}) capable of more accurately describing the HOMO-LUMO gap and energy level lineup between a molecule and two leads.  Moreover, it has been repeatedly demonstrated that a spectrum of structures exist in the experiments, some of which seem to appear more frequently than others based on the relative peak heights in histograms of the conductance \cite{Xu:2003,Venkataraman:2006,Venkataraman-Nature:2006,Tsutsui:2006,Tsutsui:2009,Xiao:2004,Huang:2007,Li:2008,Haiss:2006,Haiss:2008,Haiss:2009,Mishchenko:2010,Kim:2011}. For example, recent low-temperature (4.2 K) measurements of benzene-1,4-dithiolate (BDT) showed several peaks between 10$^{-3}$$G_{0}$ and 0.5$G_{0}$, where $G_{0}$=$\frac{2e^{2}}{h}$ \cite{Kim:2011}.  Results such as these have shifted focus away from reproducing a single value of conductance towards, more generally, determining the structures responsible for the most-probable conductance values in a given experimental setup \cite{Haiss:2009}.  Taking cues from experiments, researchers on the theoretical side have recently begun calculating the conductance of an ensemble of molecular junction structures \cite{Paulsson:2009,Strange:2010,Sergueev:2010,Pontes:2011,Andrews:2008,Cao:2008,Maul:2009,Kim:2010}.  Structures are obtained using molecular dynamics simulations in which the molecular junction is evolved through mechanical elongation \cite{Paulsson:2009,Strange:2010,Sergueev:2010,Pontes:2011} or compression \cite{Sergueev:2010}, or by thermal activation \cite{Andrews:2008,Cao:2008,Maul:2009,Kim:2010}.  Valuable information about how local structural conformations ($e.g$, oligomeric gold-thiolate units \cite{Strange:2010} and tilt angle \cite{Sergueev:2010}) influence the trends in conductance has been provided by these studies.  However, environmental factors ($e.g.$, monolayer interactions, non-ideal electrode geometry) have not yet been included in these simulations, despite the fact that they are likely to influence the results \cite{Haiss:2009,Fatemi:2011}.

Balancing accuracy and computational efficiency is a major challenge for simulations of molecular junctions.  Simulations need to accurately capture the preferred bonding geometries while also incorporating environmental factors found in experiment.  Quantum mechanical (QM)-based methods, such as density functional theory (DFT), are capable of accurately resolving molecular-level bonding, but the high computational cost of QM methods may limit the system size, reduce the total number of independent statepoints, and require simplifications to the local junction environment ($e.g.$, neglecting monolayer effects, employing ideal electrode geometries, and considering single-molecule junctions only) \cite{Paulsson:2009,Strange:2010,Sergueev:2010,Pontes:2011,Velez:2010,Sen:2010,Lin:2011}.  Additionally, energy minimizations often included in DFT calculations \cite{Sergueev:2010,Pontes:2011} may produce configurations that are not likely for thermal systems.  Methods based on classical force fields have also been used to simulate molecular junctions \cite{Andrews:2008,Cao:2008,Maul:2009,Kim:2010} and related systems \cite{Pu:2008,Pu:2010,French:2011}.  Classical force field (CFF) methods ($i.e.$, molecular dynamics -- MD -- and Monte Carlo -- MC -- simulation \cite{Frenkel:2002}) are able to handle larger system sizes and more statepoints than QM methods; however, metal-molecule interfaces exhibit complex bonding with preferred bonding sites that cannot be easily captured by conventional CFF models and methods \cite{Leng:2007}.  Previous CFF-based MD simulations of molecular junctions have only considered ideal junction environments, $e.g.$, a single molecule sandwiched between perfectly flat electrode surfaces \cite{Andrews:2008,Cao:2008,Kim:2010}. This is in contrast to experimental systems, where the bridged molecule may be surrounded by other adsorbed molecules ($i.e.$, a monolayer) with electrodes that have curved geometries resulting from, $e.g.$, the rupturing of a Au nanowire (NW), as carried out in the mechanically-controllable break junction (MCBJ) experimental technique \cite{Reed:1997,Tsutsui:2006,Tsutsui:2009}.  Reactive force fields ($e.g.,$ ReaxFF) have shown promise as a compromise between QM and CFF methods \cite{Iacovella:2011}, but parameters for metal-molecule systems are still under development.  

Here, we present an updated CFF method that balances computational efficiency and accuracy.  The method builds from our previously developed hybrid MD-MC method \cite{Pu:2010} and is capable of incorporating conditions more representative of experiment than previous work.  The technique allows larger systems with multiple bridged molecules, a large number of statepoints, and more realistic environmental factors to be considered, while retaining high accuracy through the use of bonding parameters derived from DFT calculations \cite{Leng:2007}.  Additionally, we extend our prior work \cite{Pu:2010} by performing simulations within the semigrand canonical ensemble \cite{Kofke:1988}, which includes MC moves designed to improve sampling of the preferred metal-molecule bonding geometries, further differentiating our technique from previous CFF-based studies.  Using this approach, we are able to simulate important aspects of MCBJ experiments (see Figure 1), from formation of a self-assembled monolayer (SAM) onto a Au NW surface, to elongation and rupture of the NW, and finally to trapping of a small number of molecules (between 1 and 4) in a break junction.  

%
%
\begin{figure}[h!]
	\centering
	\includegraphics[width=5.0in]{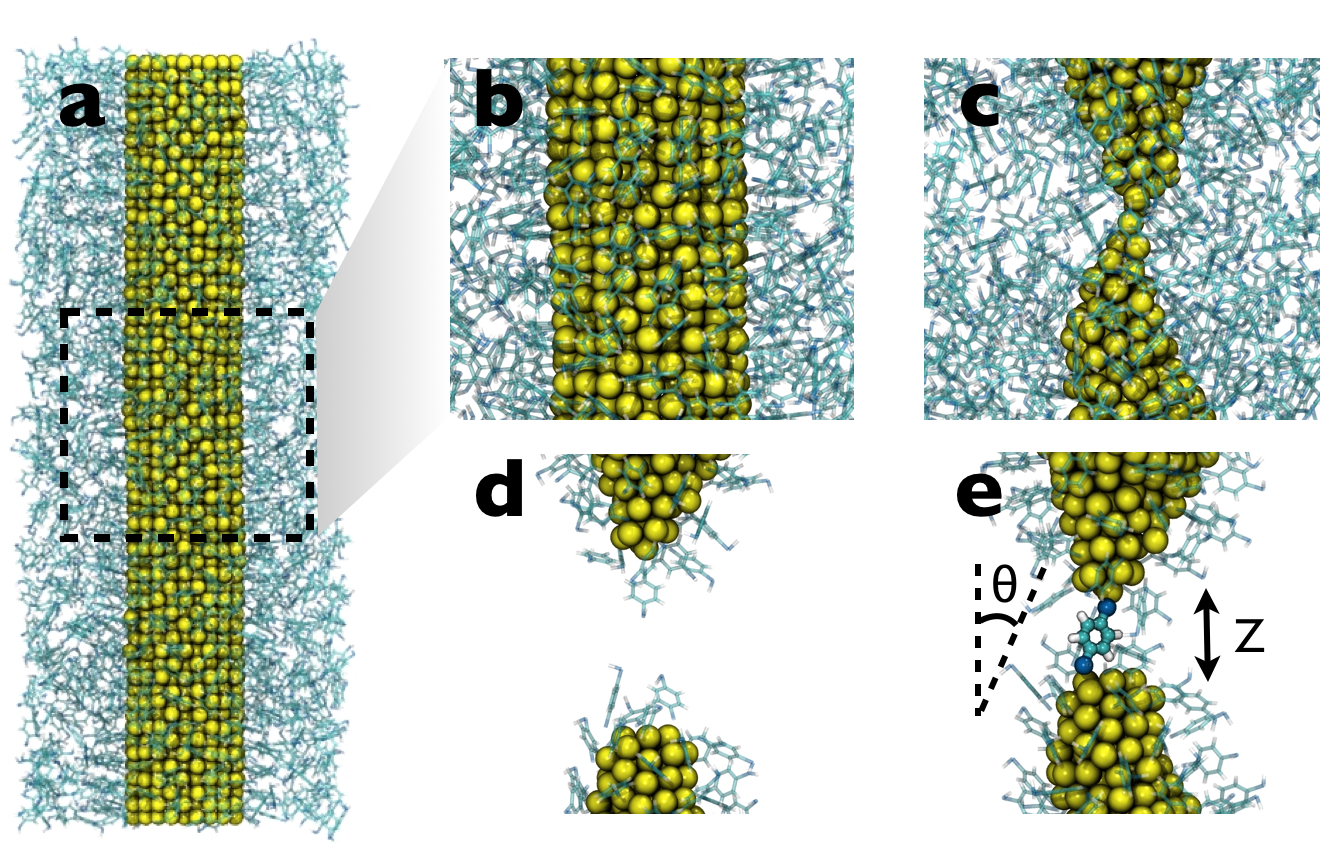}
	\caption{Simulation snapshots of the MCBJ method. (a) BDT self-assembles onto an unstretched Au NW; a closeup is shown in (b). (c) Au point contact in the necked region of the NW after $\sim$3.5 nm of elongation. (d) Following NW rupture, the bulk BDT is evaporated from the simulation box. (e) The ruptured NW tips are brought together, resulting in the formation of a molecular junction.}
	\label{fig:mcbj}
\end{figure}

We demonstrate the novelty and utility of the new simulation technique by generating a statistical ensemble of Au-BDT-Au junctions, examining in detail the number, tilt angle, and bonding geometry of bridged molecules.  By varying the extent of monolayer coverage, we find that monolayer packing is influential in the formation of single and multi-molecule junctions.  We also perform simulations with idealized tip geometries both with and without monolayers, to allow comparison with systems commonly used in simulation and theory \cite{Emberly:2001,Weber:2002,Li:2008,Mishchenko:2010,Frei:2011}. Lastly, we demonstrate that temperature can be used to control the number of bridged molecules.  The computational tractability of the simulation method allows us to perform over 1,000 simulations, resulting in statistics on par with experiment. 

\section{Results} 

To generate non-ideal electrodes that are representative of those found in MCBJ experiments, we first perform ten independent simulations of the elongation and rupture of BDT-coated 1.9-nm-diameter Au NWs (see Methods).  The NWs are elongated in the [001] direction at a rate of 1 m/s and temperature of 298 K using a hybrid MD-MC technique.  The next step in the MCBJ process, and the aspect we focus on in this paper, is the formation of a molecular junction, which we simulate using a MC-based method. Coupling each ruptured NW tip with one another (including a tip with itself) yields a total of 210 unique electrode-electrode combinations for performing simulations of the molecular junction formation process.  We simulate a mixture of two BDT species, one of which bonds at on-top sites while the other bonds at on-bridge sites.  Previous experimental \cite{Wan:2000,Li:2008} and theoretical \cite{Fischer:2003,Pontes:2006,Cossaro:2008,Leng:2007} studies have demonstrated that the on-bridge site is the energetically favored bonding site for benzenethiolate \cite{Wan:2000,Pontes:2006,Leng:2007} and alkanethiolates \cite{Fischer:2003,Cossaro:2008,Li:2008}, while on-top sites are important in low-coordination environments \cite{Li:2008}. We implement BDT identity swap moves during MC sampling, which allows BDT to chemisorb to preferred sites. The interaction potentials for BDT bonded at on-top and on-bridge sites were previously fit to DFT calculations \cite{Leng:2007}, thus enabling us to perform simulations in a classical framework while retaining high accuracy.

Following NW rupture, we displace the Au tips in the $x$-$y$ plane such that the bottom-most and top-most Au atoms in the top and bottom tips, respectively, are aligned along the $z$ axis. Next, the tips are gradually pushed together, from $Z$ = 20 \AA\ to $Z$ = 6 \AA\ (where $Z$ is the inter-electrode distance), over the course of 25 million MC moves.  We end each run at $Z$ = 6 \AA\ since direct tunneling between electrodes has been shown to occur for $Z$ < 6 \AA\ \cite{Sergueev:2010,Pontes:2011}.  Figure 2a shows a typical plot of the number of bridged BDT molecules as $Z$ is decreased.  Initially, at large values of $Z$, zero molecules are chemically attached to both electrodes.  At $Z$ < 11 \AA, a single-molecule junction forms, as shown in Figure 2b.  As $Z$ is decreased further, two (Figure 2c) and eventually three (Figure 2d) molecules connect in parallel.  All images in this paper were rendered using Visual Molecular Dynamics \cite{Humphrey:1996}.

%
%
\begin{figure}[h!]
	\centering
	\includegraphics[width=6.0in]{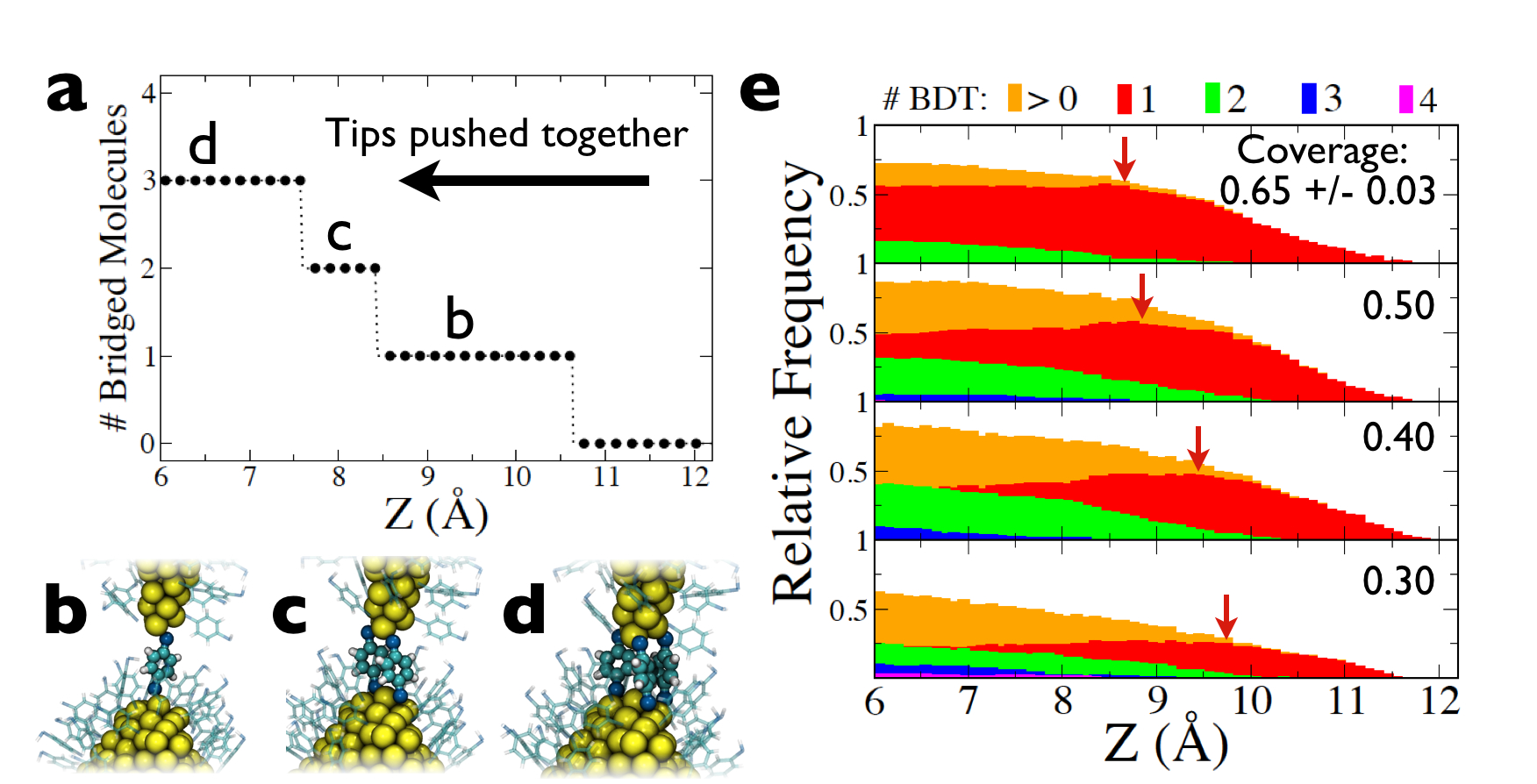}
	\caption{(a) Typical plot showing the number of bridged BDTs as the inter-electrode separation, $Z$, is decreased. This particular simulation results in (b) one bridged molecule from $Z$ $\sim$ 10.6-8.4 \AA, (c) two bridged molecules from $Z$ $\sim$ 8.4-7.6 \AA, and (d) three bridged molecules from $Z$ $\sim$ 7.6-6.0 \AA, with the corresponding images shown below.  The bridged and non-bridged BDT are rendered differently in the images for clarity. (e) Histograms of the number of bridged molecules as a function of $Z$.  The histogram bar colors correspond to the number of bridged molecules. The red arrows indicate the maximum $Z$ at which the single-molecule histograms are at least 98\%\ of their peak values.}
	\label{fig:coverage}
\end{figure}   

\subsection{Surface Coverage Effects}

We first explore the impact of monolayer packing by performing 210 simulations for each of four different surface coverages: 0.30, 0.40, 0.50, and 0.65 +/- 0.03.  Surface coverage is defined here as the number of adsorbed molecules divided by the number of Au surface atoms.  0.65 +/- 0.03 is the maximum surface coverage obtained for the 20 ruptured Au NW tips, which closely matches the reported coverage for alkanethiolates on Au nanoparticles of diameter 1.3-1.4 nm \cite{Hostetler:1998}.  

Using molecule number data such as those shown in Figure 2a, we construct histograms (see Figure 2e) of the number of bridged molecules as a function of $Z$, with separate panels representing (from top to bottom) decreasing surface coverage and the color of the histogram bars corresponding to the number of bridged BDT molecules ($n$).  We observe that the histograms of bridged molecules tend to increase with decreasing $Z$, with the exception of the $n$ = 1 case, which exhibits a peak at all four surface coverages.  These peaks, which are indicated with red arrows, appear due to the rate of formation of multi-molecule junctions exceeding that of single-molecule junctions; these peaks shift to higher $Z$ for lower surface coverages.  

We also observe in Figure 2e that $n$ depends on surface coverage.  For most $Z$, the formation of at least one bridged molecule ($n$ > 0) is most likely for surface coverage 0.50 and least likely for 0.30. The optimal surface coverage for forming a single bridged BDT ($n$ = 1) depends on $Z$; for $Z$ > 10 \AA, intermediate coverages (0.40/0.50) provide the highest probability, while for $Z$ < 9 \AA, $n$ = 1 is most probable at maximum coverage (0.65 +/- 0.03). Low surface coverages (0.30/0.40) tend to result in the highest occurrence of multi-molecule junctions.  Experimentally, conductance histograms often exhibit peaks at integer multiples ($n$) of a fundamental conductance value, with $n$ corresponding to the number of molecules in the junction \cite{Xu:2003,Xiao:2004,Tsutsui:2009,Li:2008}.  Two- and three-molecule peaks often occur in break junction experiments \cite{Xu:2003,Xiao:2004,Tsutsui:2009}; four-molecule peaks have also been observed \cite{Li:2008}.  These data match our results.  Additionally, the relative peak heights in experiment generally decrease with larger $n$, which from Figure 2e holds for most surface coverages and values of $Z$ in our simulations.  We conclude that the trends we observe in our simulations are in good agreement with experimental results, thus validating our methodology.

It is important to note that surface coverage generally varies between experimental setups, with some experiments conducted at low coverages in order to provide available bonding sites for molecular bridging \cite{Tsutsui:2006,Horiguchi:2009,Haiss:2003,Haiss:2008,Haiss:2009}, and others performed with the bridging molecules diluted in a dense matrix of non-bridging adsorbate molecules \cite{Cui:2001,Fatemi:2011}.  In the seminal work of Reed and co-workers \cite{Reed:1997}, the break junction was exposed to a solution of BDT for a long period of time, resulting in a densely packed monolayer on each of the Au nanotips.  Subsequent theoretical work \cite{Emberly:2001} suggested that the low conductance observed by Reed and co-workers could be attributed to weak electrical coupling between two overlapping BDT molecules; in other words, chemical contact between a single molecule and the two electrodes was not established, owing to the lack of available bonding sites on each nanotip.  Our results show evidence of such effects, but not to the degree that a single-molecule junction cannot form.  That is, we observe that squeezing a single molecule into an already dense monolayer is compensated by the addition of a S-Au chemical bond; however, the energetic penalty for fitting more than one molecule is often too great to overcome.  Note that the tip curvatures in our simulations may differ from those of Reed $et$ $al.$ \cite{Reed:1997}, which may influence whether a molecule is able to bridge in densely packed monolayers.  We do not consider such effects here.

In addition to changing the number of available bonding sites, the packing density of a monolayer also affects the mobility of adsorbed BDT, and thus influences whether a molecule can adopt one of the specific geometries required for bridging.  The reduced interactions between adsorbed BDTs, along with the increased availability of bonding sites, is the cause of the shifting in single-molecule peaks to larger $Z$ at lower coverages, as a second molecule can more easily bridge.  This is also the cause for the large $n$ > 0 histograms at intermediate coverages, and large multi-molecule histograms at low coverage.  It is somewhat surprising that the formation of three or four bridged BDTs is more likely at low coverage since one might expect the number of molecules on each tip to be the dominant factor in determining the number of bridged molecules \cite{Asar:2009}.  We point to the reduced monolayer interactions as the cause for this somewhat counterintuitive behavior.  We also note that in experiments conducted at low coverages, there is often evidence of multi-molecule junctions \cite{Tsutsui:2009,Haiss:2008,Haiss:2009}.  While the exact surface coverage in these experiments is unknown, Figure 2e indicates that the relative frequency at which multi-molecule junctions form will depend on $Z$ and surface coverage.

\subsection{Role of Non-Ideality}   


In order to examine the impact of realistic environmental features, we next compare our MCBJ simulation results with those for idealized systems.  First we explore the effect of using an ideal tip geometry.  The ideal tip, which is shown in Figure 3a, is an atomically sharp, pyramidal structure, reminiscent of the electrode geometries used in numerous previous theoretical studies \cite{Emberly:2001,Weber:2002,Li:2008,Mishchenko:2010,Frei:2011}.  Note that for the remainder of this paper we will only consider intermediate surface coverages of 0.40.

\begin{figure}[h!]
	\centering
	\includegraphics[width=5.0in]{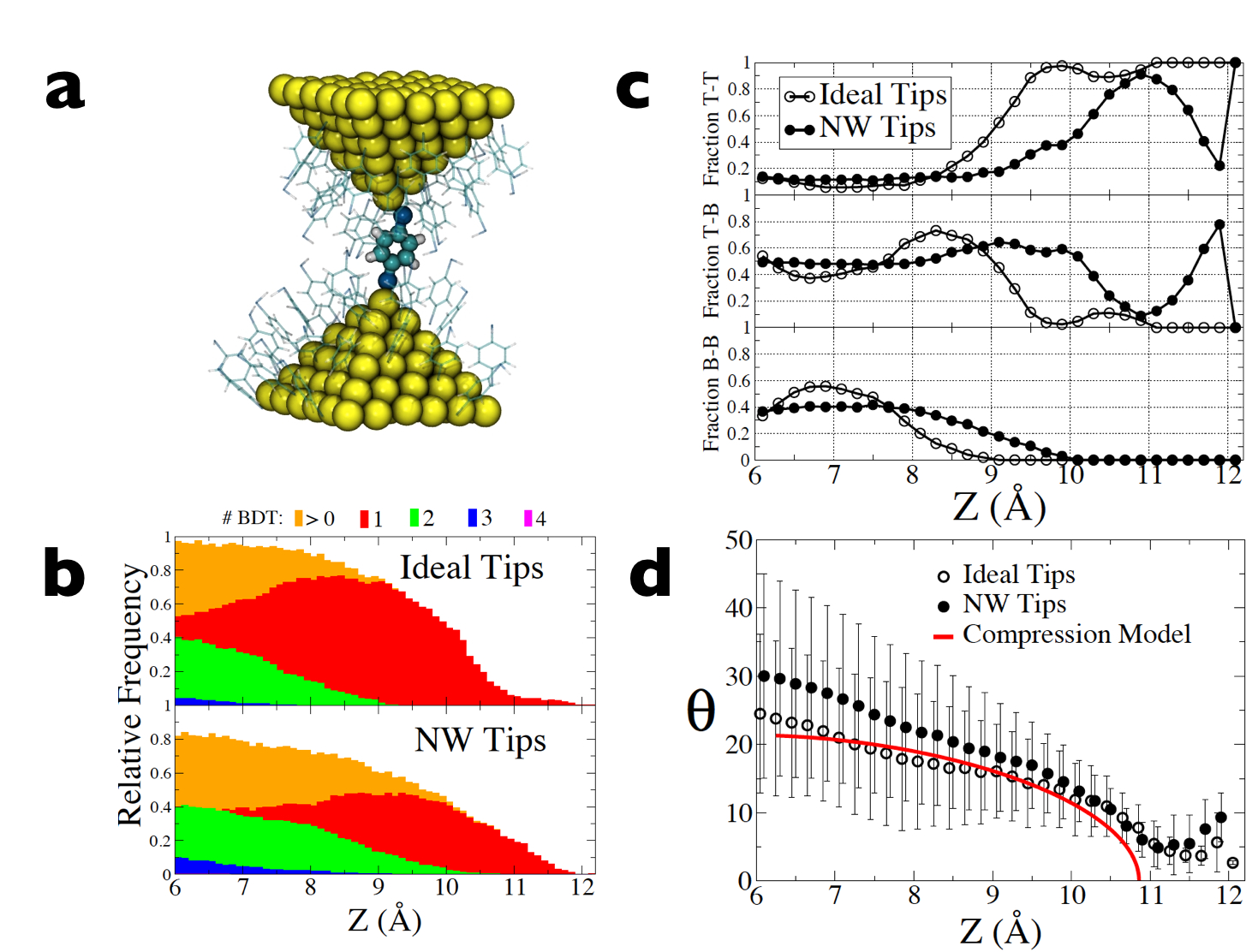}
	\caption{ (a) Molecular junction composed of two ideal, atomically sharp Au tips (at $Z$ = 10.15 \AA) and a single bridged BDT molecule.  The bridged BDT and non-bridged BDTs are rendered differently for clarity. (b) Histograms of the number of bridged molecules at various values of $Z$. (c) The bonding geometry for bridged BDT molecules as a function of $Z$.  Each panel represents the fraction of different combinations of on-top and on-bridge bonding, with (from top to bottom) ``T-T'' denoting on-top bonding at both tips, ``T-B'' denoting on-top bonding at one tip and on-bridge bonding at the other, and ``B-B'' denoting on-bridge bonding at both tips.  (d) The tilt angle, $\theta$, $versus$ $Z$. }
	\label{fig:tip-effects}
\end{figure}

Figure 3b plots histograms of the number of bridged BDT molecules as a function of $Z$, with the ideal and NW tip results shown at top and bottom, respectively. The histograms demonstrate a tip geometry dependence; the probability of having $n$ > 0 is higher for the ideal tips at $Z$ < 10 \AA, while the ideal tip histograms change more rapidly than those for the ruptured NW tips.  We further assess the impact of tip geometry by plotting the bonding geometry as a function of $Z$, shown in Figure 3c.  The separate panels display the three possible combinations of sites ($i.e.$, on-top/on-top, on-top/on-bridge, and on-bridge/on-bridge) binding a bridged molecule.  In general, on-bridge sites become more available for molecular bridging at lower values of $Z$, especially for the ideal tips where only on-top sites are accessible for bridging at high $Z$.  In contrast to the ideal tip, a ruptured NW tip can be relatively flat at its apex, with on-bridge sites accessible at high $Z$.  Lastly, we plot the tilt angle of bridged molecules, shown in Figure 3d.  Before discussing these results, we introduce a simple compression model as a first approximation for relating the inter-electrode separation, $Z$, to the tilt angle, $\theta$:

\begin{equation}Z(\theta) = D_{S-S}cos\theta + 2\sqrt{D_{S-Au}^{2}-D_{S-S}^{2}sin^{2}\theta},\label{compress}
\end{equation} where $D_{S-S}$ is the distance between S atoms in a BDT molecule (6.28 \AA\ for our rigid model of BDT) and $D_{S-Au}$ is the equilibrium S-Au bond distance (2.29 \AA\ for on-top bonding).  This model assumes that the S atoms remain bonded to the on-top sites of each tip (with $D_{S-Au}$ = 2.29 \AA), the BDT center-of-mass falls along the $z$ axis made by the two Au tips, and the tips are aligned in the $x$-$y$ plane.  We find that these first two assumptions often break down for low $Z$; nonetheless, eq. 1 establishes a baseline for comparison of tilt angle data, and qualitatively captures the behavior expected from a bridged molecule that remains at the tip apex while compressed, as opposed to one that migrates to sites along the side of a tip.  We note reasonable agreement between our compression model and the tilt angle data in Figure 3d, especially for the ideal tips.  While the tilt angle trajectory of any single bridged molecule may differ significantly from the average behavior, as evidenced by the large uncertainty bars, the average trends are in qualitative agreement with the compression model, suggesting that molecules tend not to migrate to sites along the sides of the tips.  For $Z$ < 10 \AA\ the non-ideal tips result in tilt angles that are, on average, higher than those for ideal tips, indicating that the migration of bridged molecules to sites along the side of the tips is less common in systems with non-ideal tips.


We next demonstrate the effect of a monolayer on the bonding geometry and tilt angle of bridged molecules.  After obtaining twenty different monolayer arrangements on the ideal tip (Figure 3a), we perform 210 simulations using each unique combination of the twenty BDT-decorated tips.  We identify twelve runs resulting in the formation of a single-molecule junction at $Z$ > 11 \AA.  Using these single-molecule structures as the starting point, we then perform simulations in which the remaining monolayer molecules are absent from the electrodes, enabling us to assess the impact of adsorbate interactions with the bridged molecule. 

In Figure 4a we present the bonding geometry of the bridged molecules, with similar trends observed for the monolayer and no-monolayer scenarios, but quantitative differences.  Recall that high monolayer density limits the availability of bonding sites while also reducing molecular mobility, which is responsible for the larger on-bridge peak in the no-monolayer systems shown in Figure 4a.  To further investigate why the bonding geometry changes with $Z$, it is instructive to analyze the S-Au bond energy.  In Figure 4b we present the average S-Au bond energy $versus$ $Z$ for the no-monolayer runs.  Because there is no monolayer present, the bridged molecule is able to freely explore the energetically favored sites at each tip.  At large $Z$, molecular bridging is only possible with on-top/on-top bonding geometry; as $Z$ is decreased, the energetically more stable on-bridge sites become accessible for bridging; for low values of $Z$, the compression of the tips gives rise to situations where on-bridge/on-top bonding geometry becomes energetically competitive with a somewhat strained on-bridge/on-bridge connection. 

\begin{figure}[h!]
	\centering
	\includegraphics[width=3.0in]{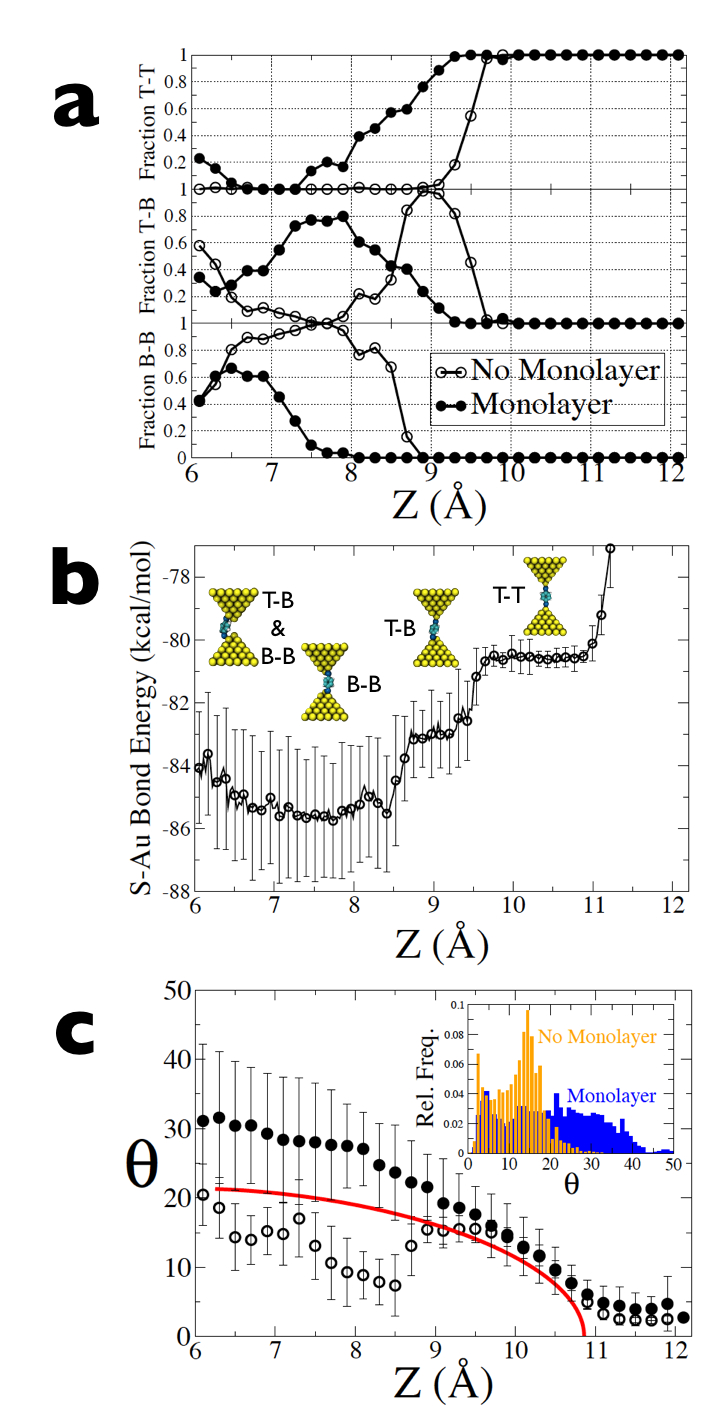}
	\caption{ Comparison of molecular break junction behavior in presence and absence of a monolayer. (a) The bonding geometry for bridged BDT molecules as a function of $Z$.  See the caption in Figure 3 for definitions of the abbreviated terms. (b) S-Au bond energy $versus$ $Z$ for a single bridged BDT molecule, neglecting monolayer effects. (c) The tilt angle, $\theta$, as a function of $Z$.  The compression model (eq. 1) is plotted as the red curve.  The inset histograms show the distribution of tilt angles. }
	\label{fig:monolayer-effects}
\end{figure}    

Figure 4c plots the tilt angle $versus$ $Z$ of bridged molecules in the presence and absence of a monolayer.  The compression model (eq. 1) is also shown (red curve) for reference.  We find that the tilt angles of bridged BDTs for $Z$ > 9.5 \AA\ agree closely for the cases where a monolayer is present and absent.  This regime is characterized by low tilt angles and, for $Z$ > 11 \AA, long S-Au bond lengths.  The maximum value of $Z$ for which a bridged molecule forms is 12.11 \AA.  This value of $Z$ requires an average S-Au bond length of 2.92 \AA, in close agreement with the reported S-Au bond rupture distance of 2.86 \AA \cite{Pontes:2011}.  For $Z$ < 9.5 \AA, the monolayer and no-monolayer results differ markedly.  In the presence of a monolayer the tilt angles of bridged molecules trend upward, indicative of the confinement of bridged molecules to the tip apex.  In absence of a monolayer, the bridged molecules exhibit different tilt angle behavior, undergoing abrupt changes that coincide with changes in the bonding geometry (see Figure 4a).  The inset in Figure 4c shows the entire distribution of tilt angles of bridged molecules.  Bridged molecules reach a maximum of $\sim$30$\degree$ in absence of a monolayer, and exhibit two preferred tilt angles at 2.5$\degree$ and 14.5$\degree$. On the other hand, the tilt angle distribution for bridged molecules in the presence of a monolayer is relatively flat from $\theta$=2-35$\degree$, with a maximum value of $\sim$50$\degree$.

The differences we highlight for idealized systems are significant since the bonding geometry and tilt angle of bridged molecules have been demonstrated to affect experimentally observed properties, namely conductance and inelastic electron tunneling spectra (IETS).  Conductance has been shown to scale linearly with the number of bridged molecules, \cite{Kushmerick:2003} while various studies \cite{Bratkovsky:2003,Tsutsui:2006,Haiss:2008,Kim:2011} have demonstrated that bonding geometry and tilt angle can affect conductance by an order of magnitude or more. For example, Haiss and co-workers \cite{Haiss:2008} showed that increasing the BDT tilt angle from $\theta$ = 0$\degree$ to $\theta$ = 50$\degree$ results in close to an order of magnitude increase in conductance, with the most pronounced increases occurring between $\theta$ = 30-50$\degree$.  Recall from the histograms in Figure 4c that the maximum tilt angle with a monolayer present is $\sim$50$\degree$, but only $\sim$30$\degree$ with no monolayer.  Thus, in this case, neglecting monolayer effects could result in significant underpredictions of conductance.  In addition to affecting conductance, bonding geometry and tilt angle have also been shown to influence the IETS of molecular junctions \cite{Lin:2011}. 

\subsection{Role of Temperature}

\begin{figure}[h!]
	\centering
	\includegraphics[width=3.0in]{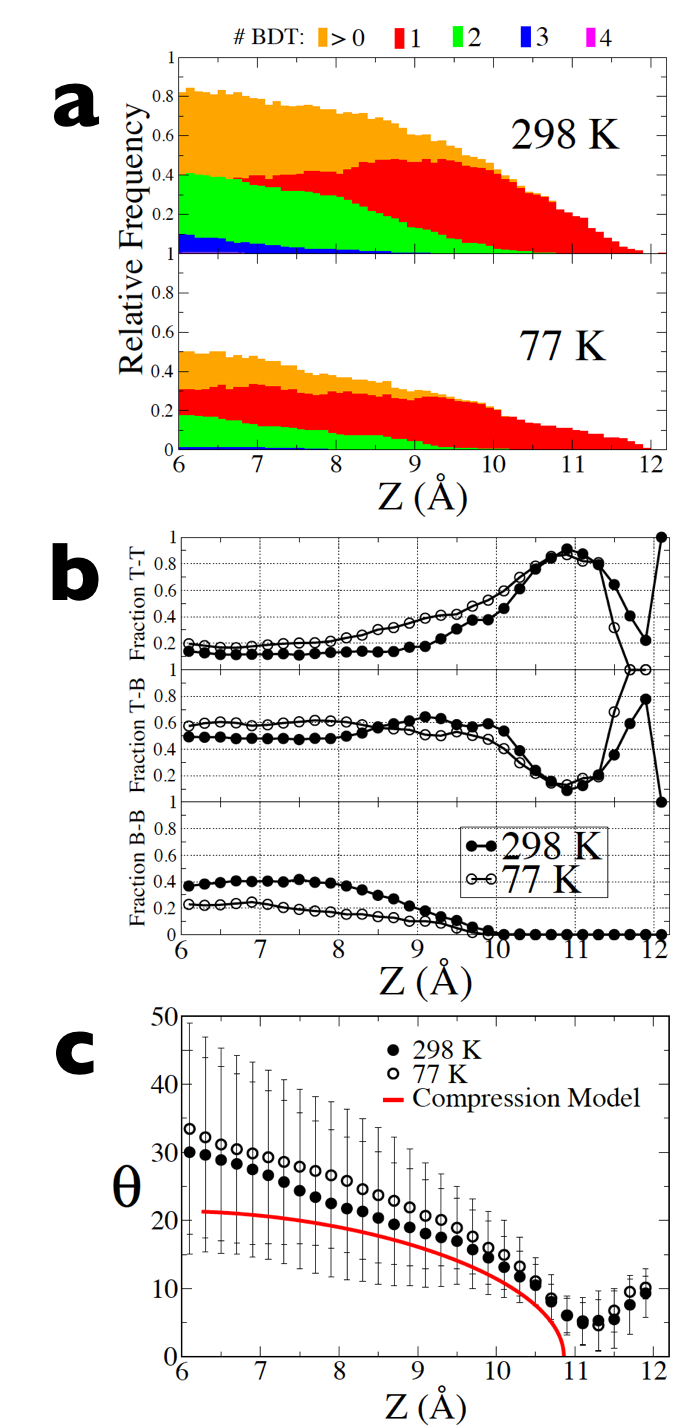}
	\caption{ Comparison of molecular break junction behavior at 298 and 77 K. (a) Histograms of the number of bridged molecules at various values of $Z$. (b) The bonding geometry for bridged BDT molecules $versus$ $Z$.  See the caption in Figure 3 for definitions of the abbreviated terms. (c) The tilt angle, $\theta$, as a function of $Z$. }
	\label{fig:temperature-effects}
\end{figure}

The results presented until now have been for a temperature of 298 K.  We now consider a temperature of 77 K, which corresponds to cryogenic conditions and has been used in experiments of Au-BDT-Au junctions \cite{Tsutsui:2009}.  We employ the same twenty ruptured NW tips for both temperatures, performing 210 simulations in each case.   Figure 5a shows histograms of the number of bridged molecules as a function of $Z$, at 298 and 77 K.  Clearly, 298 K results in a significantly higher probability of forming a molecular junction composed of any number of molecules, for a majority of $Z$; thus, the reduced mobility of the BDT molecules at 77 K is detrimental to molecular bridging. We further examine the influence of temperature by plotting the bonding geometry and tilt angle in Figure 5b and Figure 5c, respectively.  Overall, the quantitative differences between the two temperatures are small.  In Figure 5b, the fraction of on-top sites binding a bridged molecule is slightly higher at 77 K and low $Z$.  We attribute this to molecules being unable to migrate off of on-top sites after bridging there at large $Z$.  This explanation is corroborated by data in Figure 5c, which displays higher tilt angles at 77 K than 298 K, indicating that confinement of bridged molecules to the tip apex takes place more often at the lower temperature.

To our knowledge, no experimental data has been reported comparing the number, bonding geometry, or tilt angle of bridged molecules at different temperatures.  Studies on temperature-dependent behavior have focused on other properties such as mechanical stability \cite{Tsutsui:2009} and conductance \cite{Selzer:2004,Chen:2006}.  MCBJ studies of Au-BDT-Au junctions at 77 K\cite{Tsutsui:2009} and 4.2 K\cite{Kim:2011} have shown discernible peaks in histograms of conductance, but in neither case was an analysis of the relative peak heights at different temperatures reported.  At 77 K, Tsutsui and co-workers \cite{Tsutsui:2009} observed a peak in the BDT conductance histogram at 0.011$G_{0}$, matching the reported value at 298 K \cite{Tsutsui:2006,Xiao:2004}; this finding not only implies that coherent tunneling remains the dominant electron transport mechanism in the temperature range, but also that the most frequently occurring structures at 77 K and 298 K are similar.  Our results in Figure 5b and Figure 5c support this latter conclusion, especially for high $Z$, as the bonding geometry and tilt angle are very similar at the two temperatures.

\section{Discussion}

Though the precise causes are not fully understood, it is generally agreed that the environmental factors of a given experimental setup affect the conductance through a molecule.  The conductance of a bridged molecule diluted in a monolayer of non-bridging adsorbate molecules has been shown to change when different adsorbate molecules are employed \cite{Fatemi:2011}; this result was explained by changes in the relative surface coverage for different adsorbates, which can alter the electrode work function.  Building from this body of work, our results suggest that changes in the electrode work function may not be the only factor affecting conductance, as the bonding geometry and tilt angle of bridged molecules are both influenced by monolayer density.  In particular, monolayer density influences whether a molecule is able to sample the specific geometries required for bridging while also affecting the availability of preferred adsorption sites.  The detailed atomic structure of the electrodes also influences the availability of bonding sites.  Note that electrode geometry and bonding geometry have been linked previously \cite{Haiss:2009}.  Haiss $et$ $al.$ \cite{Haiss:2009} performed measurements of single-molecule conductance using four different experimental techniques, with each method producing different relative populations of the conductance histogram peaks.  The authors ascribed this behavior to changes in the electrode geometry between methods, which affected the most probable bonding geometries.  Our results also indicate that electrode geometry affects the most probable bonding geometries.  While ideal, atomically sharp tips predict a predominance of on-top/on-top bonding geometry at high $Z$, ruptured NW tip geometries allow on-bridge bonding at high $Z$.      

However, environmental factors do not greatly affect the properties of a junction in all cases.  For instance, with BDT it may be reasonable to ignore certain environmental effects for inter-electrode separations of greater than $\sim$9.5 \AA.  In this regime, the tilt angle is similar regardless of the temperature and whether adsorbate molecules are present, and the probability of forming a multi-molecule junction is low; this makes $Z$ > 9.5 \AA\ well-suited for comparisons between experiment, theory, and simulation, since simplified treatments of the junction environment do not significantly affect the properties.  On the other hand, for $Z$ < 9.5 \AA, tilt angle data diverge in cases where monolayer effects are ignored and the probability of forming multi-molecule junctions increases appreciably.  In this regime, using simplified treatments for the junction environment may result in inaccurate predictions of structure, and thus give rise to incorrect conductance results.  In this case it is necessary to perform simulations, such as those we describe in this paper, to provide input or guidance for determining the most probable structures for theoretical calculations.  

Although our simulations more closely resemble the MCBJ experimental technique than previous simulation studies, we note a few important differences between our method and the experiments.  The first difference is that unlike MCBJ experiments, the electrodes in our simulations are not contacted prior to forming each molecular junction.  While our simulations often result in molecular junctions immediately following NW rupture, for the purposes of gathering meaningful statistics and reducing computational expense, we have chosen to simulate the spontaneous formation of molecular junctions without contacting the electrodes.  In this respect, the junction formation process in our simulations is more similar to that of the $I(s)$ and $I(t)$ experimental methods of Haiss and co-workers \cite{Haiss:2006,Haiss:2008,Haiss:2009}.  We note that contacting the electrodes may help overcome activation barriers involved in junction formation, especially at lower temperature, where the spontaneous formation of a molecular junction is less likely (see Figure 5a). 

Another important difference is that while we simulate the compression of a junction, in MCBJ experiments the conductance is typically monitored as a junction is elongated.  This is an important difference considering the strength of the S-Au bond is high enough to pull short monatomic chains of Au atoms out of a surface during elongation \cite{Kruger:2002,Kim-acs-nano:2011}, and thus may result in different electrode structures than those we use.  Despite not considering such effects, we argue that the compression of a junction prior to electrode contact is a fundamental aspect of the experiments that is likely to influence the structures emerging during elongation.  Therefore, investigating the details of the compression process is essential to understanding the behavior of molecular junctions.  Furthermore, for the results presented here, we fix the structure of the Au tips during the compression/bridging portion of our simulation methodology.  This appears to be a reasonable assumption, as we did not observe significant rearrangements of the tips during test calculations that allowed the tip structure to change during compression/bridging.  However, it is important to note that experiments typically span significantly longer timescales than accessible to simulation, thus atomic rearrangements of the Au tips may additionally be important. 

Finally, we address the impact of elongation rate.  We perform simulations with a fixed elongation rate of 1 m/s and temperature of 298 K.  The rate of elongation will influence the resulting structural evolution of the wire, however, this effect will be much more significant at low temperature ($e.g.$, $\sim$4 K), in accordance with the universal energy release mechanism \cite{Pu:2008,Iacovella_scidac:2011}.  Moreover, in our previous simulations of Au NWs elongated at 298 K in vacuum, we did not observe significant differences in the spectrum of resulting tip geometries for rates ranging from 0.033 to 2 m/s, although subtle differences in the elongation pathway were observed \cite{Iacovella_scidac:2011}.  We have chosen 1 m/s for this study as it is more computationally tractable for systems including BDT.

\section{Conclusion}

We have demonstrated the utility of a new simulation method that allows for the incorporation of important environmental factors into simulations of the formation of molecular junctions.  Using this tool, we studied aspects of molecular junctions previously inaccessible with simulation.  We showed that the extent of surface coverage affects the number of bridged molecules.  Single-molecule junctions were found to occur commonly at intermediate to high surface coverages; however, at low inter-electrode separations maximum surface coverage was found to provide the highest probability of yielding single-molecule junctions, owing to the limited occurrence of multi-molecule junctions in densely packed monolayers.  We found that the reduced adsorbate-adsorbate interactions at low to intermediate surface coverages leads to relatively high probabilities for forming multi-molecule junctions.  We demonstrated that electrode geometry affects the number, bonding geometry, and tilt angle of bridged molecules.  In addition to influencing the number of bridged molecules, monolayer interactions were found to give rise to bonding geometry that is higher in energy than the preferred bonding geometry and tilt angles that are higher than those of bridged molecules in absence of a monolayer.  These are important findings since it has been previously demonstrated that both bonding geometry and tilt angle can affect conductance by at least an order of magnitude \cite{Tsutsui:2006,Kim:2011,Bratkovsky:2003,Haiss:2008}, while also impacting the measured IETS \cite{Lin:2011}.  Finally, we demonstrated that a low temperature (77 K) significantly reduces the number of bridged molecules, while resulting in only small changes in the bonding geometry and tilt angle, in comparison to 298 K.  Our results offer guidance on the design of monolayers and electrode geometries to yield desired properties, such as specific bonding geometries and/or tilt angles to control conductance.  While in this paper the focus was on Au-BDT systems, we emphasize that the simulation methodology is in principle applicable to any metal-molecule system.  

\section{Methods}

\subsection{Elongation and Rupture of BDT-Coated Au NWs}

In the hybrid MD-MC scheme, two processes are simulated that are not part of a typical MD simulation: elongation of a Au NW and chemisorption of BDT molecules onto a Au NW surface.  We implement Au NW elongation in the presence of BDT by performing MD simulations within the LAMMPS simulation package \cite{Plimpton:1995} in the canonical ensemble (constant $NVT$) with the Nos\'{e}-Hoover thermostat applied to control the temperature at 298 K.  The equations of motion are integrated using the velocity Verlet algorithm and rRESPA multi-timescale integrator, with outer and inner loop time steps of 2.0 and 0.4 fs, respectively.  Elongation is carried out using a stretch-and-relax technique in which two layers of rigid ``gripping'' atoms on one end of the wire are periodically displaced a small amount (0.1 \AA) in the [001] direction.  Two additional layers of rigid ``gripping'' atoms reside on the opposite end of the wire, while atoms within the core of the wire are dynamic.  We allow the core atoms to relax for 10 ps between displacements of the ``gripping'' atoms, corresponding to a nominal elongation rate of 1.0 m/s.  Au-Au interactions are modeled using the second-moment approximation to the tight-binding potential (TB-SMA) \cite{Cleri:1993}.  TB-SMA is a semi-empirical many-body potential capable of capturing the band character of metallic elements at a relatively low computational cost.  Furthermore, when compared to DFT calculations, TB-SMA outperforms other many-body potentials at describing the energetic and structural evolution of elongating Au NWs \cite{Pu:2007}.  BDT bonding to the Au NW is held fixed during MD runs; that is, the site at which a BDT molecule is bonded to the Au NW does not change.

In order to model BDT chemisorption, every ten elongations ($i.e.$, every 1 \AA\ of NW elongation) we perform MC sampling, with 60,000 fixed-$\mu$$VT$ (where $\mu$ is the chemical potential of a BDT molecule, $V$ is volume, and $T$ is temperature) moves followed by 160,000 fixed-$NVT$ (where $N$ is the number of BDT molecules) moves.  We find that applying this MC protocol is sufficient for equilibration of the metal-molecule interface.  That is, applying more MC moves and/or applying MC moves at more frequent elongation intervals does not change the results significantly.  The computational cost is also reasonable with this MC sampling frequency.  

The purpose of the MC simulations is to allow the BDT to efficiently sample favorable bonding sites on the Au NW surface.  Additionally, since the number of BDT molecules is allowed to change during fixed-$\mu$$VT$ MC moves, the density of BDT in the simulation box remains relatively constant for the duration of the elongation/rupture process.  During fixed-$\mu$$VT$ MC moves, we select a given move type with probabilities 0.45, 0.45, 0.04, 0.04, and 0.02 for BDT center-of-mass (COM) displacement, COM rotation, insertion, deletion, and identity swap, respectively.  For fixed-$NVT$ MC moves we select move types with probabilities 0.49, 0.49, and 0.02 for BDT COM displacement, COM rotation, and identity swap, respectively.  In accordance with the previous work of Pu $et$ $al.$ \cite{Pu-propane:2007} the excess chemical potential, $\mu_{ex}$, of both BDT species is set to -0.525 kcal/mol.

We generate the initial wire configuration by taking a cylindrical cut from a FCC lattice, oriented along the [001] direction. The NW contains a total of 2070 Au atoms, and is long enough (12.3 nm) to avoid boundary effects.  The box dimensions in the $x$ and $y$ directions are set to 5 nm, while the length of the box in the $z$ direction is variable.  We apply periodic boundary conditions in all three directions.

\subsection{Formation of Molecular Junctions}

Following NW rupture, each BDT-functionalized gold tip is allowed to relax its structure at 298 K using MD.  Since molecular junctions form locally in the break junction created by NW rupture, we extract 100 Au atoms at each tip prior to pushing the tips together, which considerably reduces the computational rigor of the simulations.  The target surface coverage is obtained by performing MC simulations at constant $\mu$$VT$.  To obtain different monolayer arrangements on the tips, we remove all chemisorbed BDT and perform the simulations with the bare Au tip as the starting point, initializing the psuedorandom number generator of each simulation with a different random seed. Next, the bulk BDT is ``evaporated'' from the simulation box, which is a standard practice in real experiments.  In the simulation, ``evaporation'' is accomplished simply by removing from the simulation box all of the BDT molecules not bonded to one or more Au atoms.  Following this, the BDT SAM is equilibrated at constant $NVT$ for 20 million MC moves.  Only one S atom in a BDT molecule is allowed to bond to the electrode during this process; however, BDT molecules are allowed to lie flat on an electrode with both S atoms bonded during the subsequent molecular junction formation runs.  In all cases the maximum BDT displacement and rotation is adjusted to obtain a 40\% acceptance rate.  The Au atoms are held fixed while the BDT molecules are modeled as rigid molecules from an optimized structure using the universal force field \cite{Rappe:1992}.  Periodic boundary conditions are applied in the $x$ and $y$ directions, while reflective walls are placed parallel to the $x$-$y$ plane at the +$z$ and -$z$ boundaries.  We use a box size of 3.5 nm in the $x$ and $y$ directions.  

The spontaneous formation of a molecular junction at fixed inter-electrode distance occurs on time scales of $\sim$0.1 s in experiment \cite{Haiss:2008}; this includes time required for bond formation and for the molecule to explore sufficient phase space for bridging.  Unfortunately, these time scales are inaccessible with molecular dynamics simulations, where time steps for integrating the equations of motion are typically on the order of 10$^{-15}$ s.  To access time scales of experiments, we model the formation of a S-Au bond using a bonding cutoff, such that if a S atom moves within 3.66 \AA\ of the appropriate bonding site (on-top or on-bridge, depending on the identity of the molecule), the S-H bond dissociates (with the H atom discarded from the simulation) and the S atom covalently bonds to the Au site.  The MC method does not provide information about the dynamics of bond formation, but rather produces thermodynamically favored, equilibrium configurations. We expect this approximation to be valid in the limit of slow compression rates, where experimental systems are given sufficient time to reach equilibrium.  We note that this treatment may slightly overpredict the formation of molecular junctions since we do not consider the details of the BDT chemisorption process, which is beyond the scope of conventional CFF methods; however, we expect the trends to be qualitatively valid, as our treatment of the bonding process is consistent for all systems.

%
%
%
\bibliography{acs-nano_library}

\providecommand*{\mcitethebibliography}{\thebibliography}
\csname @ifundefined\endcsname{endmcitethebibliography}
{\let\endmcitethebibliography\endthebibliography}{}
\begin{mcitethebibliography}{67}
\providecommand*{\natexlab}[1]{#1}
\providecommand*{\mciteSetBstSublistMode}[1]{}
\providecommand*{\mciteSetBstMaxWidthForm}[2]{}
\providecommand*{\mciteBstWouldAddEndPuncttrue}
  {\def\EndOfBibitem{\unskip.}}
\providecommand*{\mciteBstWouldAddEndPunctfalse}
  {\let\EndOfBibitem\relax}
\providecommand*{\mciteSetBstMidEndSepPunct}[3]{}
\providecommand*{\mciteSetBstSublistLabelBeginEnd}[3]{}
\providecommand*{\EndOfBibitem}{}
\mciteSetBstSublistMode{f}
\mciteSetBstMaxWidthForm{subitem}{(\alph{mcitesubitemcount})}
\mciteSetBstSublistLabelBeginEnd{\mcitemaxwidthsubitemform\space}
{\relax}{\relax}

\bibitem[Tao({2006})]{Tao:2006}
Tao,~N.~J. {Electron Transport in Molecular Junctions}. \emph{{Nat.
  Nanotechnol.}} \textbf{{2006}}, \emph{{1}}, {173--181}\relax
\mciteBstWouldAddEndPuncttrue
\mciteSetBstMidEndSepPunct{\mcitedefaultmidpunct}
{\mcitedefaultendpunct}{\mcitedefaultseppunct}\relax
\EndOfBibitem
\bibitem[Nitzan and Ratner({2003})]{Nitzan:2003}
Nitzan,~A.; Ratner,~M. {Electron Transport in Molecular Wire Junctions}.
  \emph{{Science}} \textbf{{2003}}, \emph{{300}}, {1384--1389}\relax
\mciteBstWouldAddEndPuncttrue
\mciteSetBstMidEndSepPunct{\mcitedefaultmidpunct}
{\mcitedefaultendpunct}{\mcitedefaultseppunct}\relax
\EndOfBibitem
\bibitem[Nichols et~al.({2010})Nichols, Haiss, Higgins, Leary, Martin, and
  Bethell]{Nichols:2010}
Nichols,~R.~J.; Haiss,~W.; Higgins,~S.~J.; Leary,~E.; Martin,~S.; Bethell,~D.
  {The Experimental Determination of the Conductance of Single Molecules}.
  \emph{{Phys. Chem. Chem. Phys.}} \textbf{{2010}}, \emph{{12}},
  {2801--2815}\relax
\mciteBstWouldAddEndPuncttrue
\mciteSetBstMidEndSepPunct{\mcitedefaultmidpunct}
{\mcitedefaultendpunct}{\mcitedefaultseppunct}\relax
\EndOfBibitem
\bibitem[Bandyopadhyay et~al.({2010})Bandyopadhyay, Pati, Sahu, Peper, and
  Fujita]{Bandyopadhyay:2010}
Bandyopadhyay,~A.; Pati,~R.; Sahu,~S.; Peper,~F.; Fujita,~D. {Massively
  Parallel Computing on an Organic Molecular Layer}. \emph{{Nat. Phys.}}
  \textbf{{2010}}, \emph{{6}}, {369--375}\relax
\mciteBstWouldAddEndPuncttrue
\mciteSetBstMidEndSepPunct{\mcitedefaultmidpunct}
{\mcitedefaultendpunct}{\mcitedefaultseppunct}\relax
\EndOfBibitem
\bibitem[Reed et~al.(1997)Reed, Zhou, Muller, Burgin, and Tour]{Reed:1997}
Reed,~M.~A.; Zhou,~C.; Muller,~C.~J.; Burgin,~T.~P.; Tour,~J.~M. Conductance of
  a Molecular Junction. \emph{Science} \textbf{1997}, \emph{278},
  252--254\relax
\mciteBstWouldAddEndPuncttrue
\mciteSetBstMidEndSepPunct{\mcitedefaultmidpunct}
{\mcitedefaultendpunct}{\mcitedefaultseppunct}\relax
\EndOfBibitem
\bibitem[Di~Ventra et~al.({2000})Di~Ventra, Pantelides, and
  Lang]{DiVentra:2000}
Di~Ventra,~M.; Pantelides,~S.; Lang,~N. {First-Principles Calculation of
  Transport Properties of a Molecular Device}. \emph{{Phys. Rev. Lett.}}
  \textbf{{2000}}, \emph{{84}}, {979--982}\relax
\mciteBstWouldAddEndPuncttrue
\mciteSetBstMidEndSepPunct{\mcitedefaultmidpunct}
{\mcitedefaultendpunct}{\mcitedefaultseppunct}\relax
\EndOfBibitem
\bibitem[Lindsay and Ratner({2007})]{Lindsay:2007}
Lindsay,~S.~M.; Ratner,~M.~A. {Molecular Transport Junctions: Clearing Mists}.
  \emph{{Adv. Mat.}} \textbf{{2007}}, \emph{{19}}, {23--31}\relax
\mciteBstWouldAddEndPuncttrue
\mciteSetBstMidEndSepPunct{\mcitedefaultmidpunct}
{\mcitedefaultendpunct}{\mcitedefaultseppunct}\relax
\EndOfBibitem
\bibitem[Xu and Tao({2003})]{Xu:2003}
Xu,~B.; Tao,~N. {Measurement of Single-Molecule Resistance by Repeated
  Formation of Molecular Junctions}. \emph{{Science}} \textbf{{2003}},
  \emph{{301}}, {1221--1223}\relax
\mciteBstWouldAddEndPuncttrue
\mciteSetBstMidEndSepPunct{\mcitedefaultmidpunct}
{\mcitedefaultendpunct}{\mcitedefaultseppunct}\relax
\EndOfBibitem
\bibitem[Venkataraman et~al.({2006})Venkataraman, Klare, Tam, Nuckolls,
  Hybertsen, and Steigerwald]{Venkataraman:2006}
Venkataraman,~L.; Klare,~J.; Tam,~I.; Nuckolls,~C.; Hybertsen,~M.;
  Steigerwald,~M. {Single-Molecule Circuits with Well-Defined Molecular
  Conductance}. \emph{{Nano Lett.}} \textbf{{2006}}, \emph{{6}},
  {458--462}\relax
\mciteBstWouldAddEndPuncttrue
\mciteSetBstMidEndSepPunct{\mcitedefaultmidpunct}
{\mcitedefaultendpunct}{\mcitedefaultseppunct}\relax
\EndOfBibitem
\bibitem[Venkataraman et~al.({2006})Venkataraman, Klare, Nuckolls, Hybertsen,
  and Steigerwald]{Venkataraman-Nature:2006}
Venkataraman,~L.; Klare,~J.~E.; Nuckolls,~C.; Hybertsen,~M.~S.;
  Steigerwald,~M.~L. {Dependence of Single-Molecule Junction Conductance on
  Molecular Conformation}. \emph{{Nature}} \textbf{{2006}}, \emph{{442}},
  {904--907}\relax
\mciteBstWouldAddEndPuncttrue
\mciteSetBstMidEndSepPunct{\mcitedefaultmidpunct}
{\mcitedefaultendpunct}{\mcitedefaultseppunct}\relax
\EndOfBibitem
\bibitem[Tsutsui et~al.({2006})Tsutsui, Teramae, Kurokawa, and
  Sakai]{Tsutsui:2006}
Tsutsui,~M.; Teramae,~Y.; Kurokawa,~S.; Sakai,~A. {High-Conductance States of
  Single Benzenedithiol Molecules}. \emph{{Appl. Phys. Lett.}} \textbf{{2006}},
  \emph{{89}}, {163111}\relax
\mciteBstWouldAddEndPuncttrue
\mciteSetBstMidEndSepPunct{\mcitedefaultmidpunct}
{\mcitedefaultendpunct}{\mcitedefaultseppunct}\relax
\EndOfBibitem
\bibitem[Tsutsui et~al.({2009})Tsutsui, Taniguchi, and Kawai]{Tsutsui:2009}
Tsutsui,~M.; Taniguchi,~M.; Kawai,~T. {Atomistic Mechanics and Formation
  Mechanism of Metal-Molecule-Metal Junctions}. \emph{{Nano Lett.}}
  \textbf{{2009}}, \emph{{9}}, {2433--2439}\relax
\mciteBstWouldAddEndPuncttrue
\mciteSetBstMidEndSepPunct{\mcitedefaultmidpunct}
{\mcitedefaultendpunct}{\mcitedefaultseppunct}\relax
\EndOfBibitem
\bibitem[Strange et~al.({2011})Strange, Rostgaard, Hakkinen, and
  Thygesen]{Strange:2011}
Strange,~M.; Rostgaard,~C.; Hakkinen,~H.; Thygesen,~K.~S. {Self-Consistent GW
  Calculations of Electronic Transport in Thiol- and Amine-Linked Molecular
  Junctions}. \emph{{Phys. Rev. B}} \textbf{{2011}}, \emph{{83}},
  {115108}\relax
\mciteBstWouldAddEndPuncttrue
\mciteSetBstMidEndSepPunct{\mcitedefaultmidpunct}
{\mcitedefaultendpunct}{\mcitedefaultseppunct}\relax
\EndOfBibitem
\bibitem[Toher and Sanvito({2007})]{Toher:2007}
Toher,~C.; Sanvito,~S. {Efficient Atomic Self-Interaction Correction Scheme for
  Nonequilibrium Quantum Transport}. \emph{{Phys. Rev. Lett.}} \textbf{{2007}},
  \emph{{99}}, {056801}\relax
\mciteBstWouldAddEndPuncttrue
\mciteSetBstMidEndSepPunct{\mcitedefaultmidpunct}
{\mcitedefaultendpunct}{\mcitedefaultseppunct}\relax
\EndOfBibitem
\bibitem[Pontes et~al.({2011})Pontes, Rocha, Sanvito, Fazzio, and Roque~da
  Silva]{Pontes:2011}
Pontes,~R.~B.; Rocha,~A.~R.; Sanvito,~S.; Fazzio,~A.; Roque~da Silva,~A.~J. {Ab
  Initio Calculations of Structural Evolution and Conductance of
  Benzene-1,4-dithiol on Gold Leads}. \emph{{ACS Nano}} \textbf{{2011}},
  \emph{{5}}, {795--804}\relax
\mciteBstWouldAddEndPuncttrue
\mciteSetBstMidEndSepPunct{\mcitedefaultmidpunct}
{\mcitedefaultendpunct}{\mcitedefaultseppunct}\relax
\EndOfBibitem
\bibitem[Xiao et~al.({2004})Xiao, Xu, and Tao]{Xiao:2004}
Xiao,~X.; Xu,~B.; Tao,~N. {Measurement of Single Molecule Conductance:
  Benzenedithiol and Benzenedimethanethiol}. \emph{{Nano Lett.}}
  \textbf{{2004}}, \emph{{4}}, {267--271}\relax
\mciteBstWouldAddEndPuncttrue
\mciteSetBstMidEndSepPunct{\mcitedefaultmidpunct}
{\mcitedefaultendpunct}{\mcitedefaultseppunct}\relax
\EndOfBibitem
\bibitem[Huang et~al.({2007})Huang, Chen, Bennett, and Tao]{Huang:2007}
Huang,~Z.; Chen,~F.; Bennett,~P.~A.; Tao,~N. {Single Molecule Junctions Formed
  via Au-Thiol Contact: Stability and Breakdown Mechanism}. \emph{{J. Am. Chem.
  Soc.}} \textbf{{2007}}, \emph{{129}}, {13225--13231}\relax
\mciteBstWouldAddEndPuncttrue
\mciteSetBstMidEndSepPunct{\mcitedefaultmidpunct}
{\mcitedefaultendpunct}{\mcitedefaultseppunct}\relax
\EndOfBibitem
\bibitem[Li et~al.({2008})Li, Pobelov, Wandlowski, Bagrets, Arnold, and
  Evers]{Li:2008}
Li,~C.; Pobelov,~I.; Wandlowski,~T.; Bagrets,~A.; Arnold,~A.; Evers,~F. {Charge
  Transport in Single Au Vertical Bar Alkanedithiol Vertical Bar Au Junctions:
  Coordination Geometries and Conformational Degrees of Freedom}. \emph{{J. Am.
  Chem. Soc.}} \textbf{{2008}}, \emph{{130}}, {318--326}\relax
\mciteBstWouldAddEndPuncttrue
\mciteSetBstMidEndSepPunct{\mcitedefaultmidpunct}
{\mcitedefaultendpunct}{\mcitedefaultseppunct}\relax
\EndOfBibitem
\bibitem[Haiss et~al.({2006})Haiss, Wang, Grace, Batsanov, Schiffrin, Higgins,
  Bryce, Lambert, and Nichols]{Haiss:2006}
Haiss,~W.; Wang,~C.; Grace,~I.; Batsanov,~A.~S.; Schiffrin,~D.~J.;
  Higgins,~S.~J.; Bryce,~M.~R.; Lambert,~C.~J.; Nichols,~R.~J. {Precision
  Control of Single-Molecule Electrical Junctions}. \emph{{Nat. Mat.}}
  \textbf{{2006}}, \emph{{5}}, {995--1002}\relax
\mciteBstWouldAddEndPuncttrue
\mciteSetBstMidEndSepPunct{\mcitedefaultmidpunct}
{\mcitedefaultendpunct}{\mcitedefaultseppunct}\relax
\EndOfBibitem
\bibitem[Haiss et~al.({2008})Haiss, Wang, Jitchati, Grace, Martin, Batsanov,
  Higgins, Bryce, Lambert, Jensen, and Nichols]{Haiss:2008}
Haiss,~W.; Wang,~C.; Jitchati,~R.; Grace,~I.; Martin,~S.; Batsanov,~A.~S.;
  Higgins,~S.~J.; Bryce,~M.~R.; Lambert,~C.~J.; Jensen,~P.~S. et~al. {Variable
  Contact Gap Single-Molecule Conductance Determination for a Series of
  Conjugated Molecular Bridges}. \emph{{J. Phys. Cond. Matt.}} \textbf{{2008}},
  \emph{{20}}, {374119}\relax
\mciteBstWouldAddEndPuncttrue
\mciteSetBstMidEndSepPunct{\mcitedefaultmidpunct}
{\mcitedefaultendpunct}{\mcitedefaultseppunct}\relax
\EndOfBibitem
\bibitem[Haiss et~al.({2009})Haiss, Martin, Leary, van Zalinge, Higgins,
  Bouffier, and Nichols]{Haiss:2009}
Haiss,~W.; Martin,~S.; Leary,~E.; van Zalinge,~H.; Higgins,~S.~J.;
  Bouffier,~L.; Nichols,~R.~J. {Impact of Junction Formation Method and Surface
  Roughness on Single Molecule Conductance}. \emph{{J. Phys. Chem. C}}
  \textbf{{2009}}, \emph{{113}}, {5823--5833}\relax
\mciteBstWouldAddEndPuncttrue
\mciteSetBstMidEndSepPunct{\mcitedefaultmidpunct}
{\mcitedefaultendpunct}{\mcitedefaultseppunct}\relax
\EndOfBibitem
\bibitem[Mishchenko et~al.({2010})Mishchenko, Vonlanthen, Meded, Buerkle, Li,
  Pobelov, Bagrets, Viljas, Pauly, Evers, Mayor, and
  Wandlowski]{Mishchenko:2010}
Mishchenko,~A.; Vonlanthen,~D.; Meded,~V.; Buerkle,~M.; Li,~C.; Pobelov,~I.~V.;
  Bagrets,~A.; Viljas,~J.~K.; Pauly,~F.; Evers,~F. et~al. {Influence of
  Conformation on Conductance of Biphenyl-Dithiol Single-Molecule Contacts}.
  \emph{{Nano Lett.}} \textbf{{2010}}, \emph{{10}}, {156--163}\relax
\mciteBstWouldAddEndPuncttrue
\mciteSetBstMidEndSepPunct{\mcitedefaultmidpunct}
{\mcitedefaultendpunct}{\mcitedefaultseppunct}\relax
\EndOfBibitem
\bibitem[Kim et~al.({2011})Kim, Pietsch, Erbe, Belzig, and Scheer]{Kim:2011}
Kim,~Y.; Pietsch,~T.; Erbe,~A.; Belzig,~W.; Scheer,~E. {Benzenedithiol: A
  Broad-Range Single-Channel Molecular Conductor}. \emph{{Nano Lett.}}
  \textbf{{2011}}, \emph{{11}}, {3734--3738}\relax
\mciteBstWouldAddEndPuncttrue
\mciteSetBstMidEndSepPunct{\mcitedefaultmidpunct}
{\mcitedefaultendpunct}{\mcitedefaultseppunct}\relax
\EndOfBibitem
\bibitem[Paulsson et~al.({2009})Paulsson, Krag, Frederiksen, and
  Brandbyge]{Paulsson:2009}
Paulsson,~M.; Krag,~C.; Frederiksen,~T.; Brandbyge,~M. {Conductance of
  Alkanedithiol Single-Molecule Junctions: A Molecular Dynamics Study}.
  \emph{{Nano Lett.}} \textbf{{2009}}, \emph{{9}}, {117--121}\relax
\mciteBstWouldAddEndPuncttrue
\mciteSetBstMidEndSepPunct{\mcitedefaultmidpunct}
{\mcitedefaultendpunct}{\mcitedefaultseppunct}\relax
\EndOfBibitem
\bibitem[Strange et~al.({2010})Strange, Lopez-Acevedo, and
  Hakkinen]{Strange:2010}
Strange,~M.; Lopez-Acevedo,~O.; Hakkinen,~H. {Oligomeric Gold-Thiolate Units
  Define the Properties of the Molecular Junction between Gold and Benzene
  Dithiols}. \emph{{J. Phys. Chem. Lett.}} \textbf{{2010}}, \emph{{1}},
  {1528--1532}\relax
\mciteBstWouldAddEndPuncttrue
\mciteSetBstMidEndSepPunct{\mcitedefaultmidpunct}
{\mcitedefaultendpunct}{\mcitedefaultseppunct}\relax
\EndOfBibitem
\bibitem[Sergueev et~al.({2010})Sergueev, Tsetseris, Varga, and
  Pantelides]{Sergueev:2010}
Sergueev,~N.; Tsetseris,~L.; Varga,~K.; Pantelides,~S. {Configuration and
  Conductance Evolution of Benzene-Dithiol Molecular Junctions Under
  Elongation}. \emph{{Phys. Rev. B}} \textbf{{2010}}, \emph{{82}},
  {073106}\relax
\mciteBstWouldAddEndPuncttrue
\mciteSetBstMidEndSepPunct{\mcitedefaultmidpunct}
{\mcitedefaultendpunct}{\mcitedefaultseppunct}\relax
\EndOfBibitem
\bibitem[Andrews et~al.({2008})Andrews, Van~Duyne, and Ratner]{Andrews:2008}
Andrews,~D.~Q.; Van~Duyne,~R.~P.; Ratner,~M.~A. {Stochastic Modulation in
  Molecular Electronic Transport Junctions: Molecular Dynamics Coupled with
  Charge Transport Calculations}. \emph{{Nano Lett.}} \textbf{{2008}},
  \emph{{8}}, {1120--1126}\relax
\mciteBstWouldAddEndPuncttrue
\mciteSetBstMidEndSepPunct{\mcitedefaultmidpunct}
{\mcitedefaultendpunct}{\mcitedefaultseppunct}\relax
\EndOfBibitem
\bibitem[Cao et~al.(2008)Cao, Jiang, Ma, and Luo]{Cao:2008}
Cao,~H.; Jiang,~J.; Ma,~J.; Luo,~Y. Temperature-Dependent Statistical Behavior
  of Single Molecular Conductance in Aqueous Solution. \emph{J. Am. Chem. Soc.}
  \textbf{2008}, \emph{130}, 6674--6675\relax
\mciteBstWouldAddEndPuncttrue
\mciteSetBstMidEndSepPunct{\mcitedefaultmidpunct}
{\mcitedefaultendpunct}{\mcitedefaultseppunct}\relax
\EndOfBibitem
\bibitem[Maul and Wenzel(2009)]{Maul:2009}
Maul,~R.; Wenzel,~W. Influence of structural disorder and large-scale geometric
  fluctuations on the coherent transport of metallic junctions and molecular
  wires. \emph{Phys. Rev. B} \textbf{2009}, \emph{80}, 045424\relax
\mciteBstWouldAddEndPuncttrue
\mciteSetBstMidEndSepPunct{\mcitedefaultmidpunct}
{\mcitedefaultendpunct}{\mcitedefaultseppunct}\relax
\EndOfBibitem
\bibitem[Kim and Kim({2010})]{Kim:2010}
Kim,~H.~S.; Kim,~Y.-H. {Conformational and Conductance Fluctuations in a
  Single-Molecule Junction: Multiscale Computational Study}. \emph{{Phys. Rev.
  B}} \textbf{{2010}}, \emph{{82}}, {075412}\relax
\mciteBstWouldAddEndPuncttrue
\mciteSetBstMidEndSepPunct{\mcitedefaultmidpunct}
{\mcitedefaultendpunct}{\mcitedefaultseppunct}\relax
\EndOfBibitem
\bibitem[Fatemi et~al.({2011})Fatemi, Kamenetska, Neaton, and
  Venkataraman]{Fatemi:2011}
Fatemi,~V.; Kamenetska,~M.; Neaton,~J.~B.; Venkataraman,~L. {Environmental
  Control of Single-Molecule Junction Transport}. \emph{{Nano Lett.}}
  \textbf{{2011}}, \emph{{11}}, {1988--1992}\relax
\mciteBstWouldAddEndPuncttrue
\mciteSetBstMidEndSepPunct{\mcitedefaultmidpunct}
{\mcitedefaultendpunct}{\mcitedefaultseppunct}\relax
\EndOfBibitem
\bibitem[Velez et~al.({2010})Velez, Dassie, and Leiva]{Velez:2010}
Velez,~P.; Dassie,~S.~A.; Leiva,~E. P.~M. {Role of Metal Contacts in the
  Mechanical Properties of Molecular Nanojunctions: Comparative Ab Initio Study
  of Au/1,8-Octanedithiol and Au/4,4-Bipyridine}. \emph{{Phys. Rev. B}}
  \textbf{{2010}}, \emph{{81}}, {235435}\relax
\mciteBstWouldAddEndPuncttrue
\mciteSetBstMidEndSepPunct{\mcitedefaultmidpunct}
{\mcitedefaultendpunct}{\mcitedefaultseppunct}\relax
\EndOfBibitem
\bibitem[Sen and Kaun({2010})]{Sen:2010}
Sen,~A.; Kaun,~C.-C. {Effect of Electrode Orientations on Charge Transport in
  Alkanedithiol Single-Molecule Junctions}. \emph{{ACS Nano}} \textbf{{2010}},
  \emph{{4}}, {6404--6408}\relax
\mciteBstWouldAddEndPuncttrue
\mciteSetBstMidEndSepPunct{\mcitedefaultmidpunct}
{\mcitedefaultendpunct}{\mcitedefaultseppunct}\relax
\EndOfBibitem
\bibitem[Lin et~al.({2011})Lin, Wang, and Luo]{Lin:2011}
Lin,~L.-L.; Wang,~C.-K.; Luo,~Y. {Inelastic Electron Tunneling Spectroscopy of
  Gold-Benzenedithiol-Gold Junctions: Accurate Determination of Molecular
  Conformation}. \emph{{ACS Nano}} \textbf{{2011}}, \emph{{5}},
  {2257--2263}\relax
\mciteBstWouldAddEndPuncttrue
\mciteSetBstMidEndSepPunct{\mcitedefaultmidpunct}
{\mcitedefaultendpunct}{\mcitedefaultseppunct}\relax
\EndOfBibitem
\bibitem[Pu et~al.(2008)Pu, Leng, and Cummings]{Pu:2008}
Pu,~Q.; Leng,~Y.; Cummings,~P.~T. Rate-Dependent Energy Release Mechanism of
  Gold Nanowires under Elongation. \emph{J. Am. Chem. Soc.} \textbf{2008},
  \emph{130}, 17907--17912\relax
\mciteBstWouldAddEndPuncttrue
\mciteSetBstMidEndSepPunct{\mcitedefaultmidpunct}
{\mcitedefaultendpunct}{\mcitedefaultseppunct}\relax
\EndOfBibitem
\bibitem[Pu et~al.({2010})Pu, Leng, Zhao, and Cummings]{Pu:2010}
Pu,~Q.; Leng,~Y.; Zhao,~X.; Cummings,~P.~T. {Molecular Simulation Studies on
  the Elongation of Gold Nanowires in Benzenedithiol}. \emph{{J. Phys. Chem.
  C}} \textbf{{2010}}, \emph{{114}}, {10365--10372}\relax
\mciteBstWouldAddEndPuncttrue
\mciteSetBstMidEndSepPunct{\mcitedefaultmidpunct}
{\mcitedefaultendpunct}{\mcitedefaultseppunct}\relax
\EndOfBibitem
\bibitem[French et~al.(2011)French, Iacovella, and Cummings]{French:2011}
French,~W.~R.; Iacovella,~C.~R.; Cummings,~P.~T. The Influence of Molecular
  Adsorption on Elongating Gold Nanowires. \emph{J. Phys. Chem. C}
  \textbf{2011}, \emph{115}, 18422--18433\relax
\mciteBstWouldAddEndPuncttrue
\mciteSetBstMidEndSepPunct{\mcitedefaultmidpunct}
{\mcitedefaultendpunct}{\mcitedefaultseppunct}\relax
\EndOfBibitem
\bibitem[Frenkel and Smit({2002})]{Frenkel:2002}
Frenkel,~D.; Smit,~B. \emph{{Understanding Molecular Simulation: From
  Algorithms to Applications}};
\newblock {Academic Press}: {San Diego}, {2002}\relax
\mciteBstWouldAddEndPuncttrue
\mciteSetBstMidEndSepPunct{\mcitedefaultmidpunct}
{\mcitedefaultendpunct}{\mcitedefaultseppunct}\relax
\EndOfBibitem
\bibitem[Leng et~al.({2007})Leng, Dyer, Krstic, Harrison, and
  Cummings]{Leng:2007}
Leng,~Y.~S.; Dyer,~P.~J.; Krstic,~P.~S.; Harrison,~R.~J.; Cummings,~P.~T.
  {Calibration of Chemical Bonding Between Benzenedithiolate and Gold: The
  Effects of Geometry and Size of Gold Clusters}. \emph{{Mol. Phys.}}
  \textbf{{2007}}, \emph{{105}}, {293--300}\relax
\mciteBstWouldAddEndPuncttrue
\mciteSetBstMidEndSepPunct{\mcitedefaultmidpunct}
{\mcitedefaultendpunct}{\mcitedefaultseppunct}\relax
\EndOfBibitem
\bibitem[Iacovella et~al.(2011)Iacovella, French, Cook, Kent, and
  Cummings]{Iacovella:2011}
Iacovella,~C.~R.; French,~W.~R.; Cook,~B.~G.; Kent,~P. R.~C.; Cummings,~P.~T.
  Role of Polytetrahedral Structures in the Elongation and Rupture of Gold
  Nanowires. \emph{ACS Nano} \textbf{2011}, \emph{5}, 10065--10073\relax
\mciteBstWouldAddEndPuncttrue
\mciteSetBstMidEndSepPunct{\mcitedefaultmidpunct}
{\mcitedefaultendpunct}{\mcitedefaultseppunct}\relax
\EndOfBibitem
\bibitem[Kofke and Glandt({1988})]{Kofke:1988}
Kofke,~D.; Glandt,~E. {Monte-Carlo Simulation of Multicomponent Equilibria in a
  Semigrand Canonical Ensemble}. \emph{{Mol. Phys.}} \textbf{{1988}},
  \emph{{64}}, {1105--1131}\relax
\mciteBstWouldAddEndPuncttrue
\mciteSetBstMidEndSepPunct{\mcitedefaultmidpunct}
{\mcitedefaultendpunct}{\mcitedefaultseppunct}\relax
\EndOfBibitem
\bibitem[Emberly and Kirczenow({2001})]{Emberly:2001}
Emberly,~E.; Kirczenow,~G. {Models of Electron Transport through Organic
  Molecular Monolayers Self-Assembled on Nanoscale Metallic Contacts}.
  \emph{{Phys. Rev. B}} \textbf{{2001}}, \emph{{64}}, {235412}\relax
\mciteBstWouldAddEndPuncttrue
\mciteSetBstMidEndSepPunct{\mcitedefaultmidpunct}
{\mcitedefaultendpunct}{\mcitedefaultseppunct}\relax
\EndOfBibitem
\bibitem[Weber et~al.({2002})Weber, Reichert, Weigend, Ochs, Beckmann, Mayor,
  Ahlrichs, and von Lohneysen]{Weber:2002}
Weber,~H.; Reichert,~J.; Weigend,~F.; Ochs,~R.; Beckmann,~D.; Mayor,~M.;
  Ahlrichs,~R.; von Lohneysen,~H. {Electronic Transport through Single
  Conjugated Molecules}. \emph{{Chem. Phys.}} \textbf{{2002}}, \emph{{281}},
  {113--125}\relax
\mciteBstWouldAddEndPuncttrue
\mciteSetBstMidEndSepPunct{\mcitedefaultmidpunct}
{\mcitedefaultendpunct}{\mcitedefaultseppunct}\relax
\EndOfBibitem
\bibitem[Frei et~al.({2011})Frei, Aradhya, Koentopp, Hybertsen, and
  Venkataraman]{Frei:2011}
Frei,~M.; Aradhya,~S.~V.; Koentopp,~M.; Hybertsen,~M.~S.; Venkataraman,~L.
  {Mechanics and Chemistry: Single Molecule Bond Rupture Forces Correlate with
  Molecular Backbone Structure}. \emph{{Nano Lett.}} \textbf{{2011}},
  \emph{{11}}, {1518--1523}\relax
\mciteBstWouldAddEndPuncttrue
\mciteSetBstMidEndSepPunct{\mcitedefaultmidpunct}
{\mcitedefaultendpunct}{\mcitedefaultseppunct}\relax
\EndOfBibitem
\bibitem[Wan et~al.({2000})Wan, Terashima, Noda, and Osawa]{Wan:2000}
Wan,~L.; Terashima,~M.; Noda,~H.; Osawa,~M. {Molecular Orientation and Ordered
  Structure of Benzenethiol Adsorbed on Gold(111)}. \emph{{J. Phys. Chem. B}}
  \textbf{{2000}}, \emph{{104}}, {3563--3569}\relax
\mciteBstWouldAddEndPuncttrue
\mciteSetBstMidEndSepPunct{\mcitedefaultmidpunct}
{\mcitedefaultendpunct}{\mcitedefaultseppunct}\relax
\EndOfBibitem
\bibitem[Fischer et~al.({2003})Fischer, Curioni, and Andreoni]{Fischer:2003}
Fischer,~D.; Curioni,~A.; Andreoni,~W. {Decanethiols on Gold: The Structure of
  Self-Assembled Monolayers Unraveled with Computer Simulations}.
  \emph{{Langmuir}} \textbf{{2003}}, \emph{{19}}, {3567--3571}\relax
\mciteBstWouldAddEndPuncttrue
\mciteSetBstMidEndSepPunct{\mcitedefaultmidpunct}
{\mcitedefaultendpunct}{\mcitedefaultseppunct}\relax
\EndOfBibitem
\bibitem[Pontes et~al.(2006)Pontes, Novaes, Fazzio, and da~Silva]{Pontes:2006}
Pontes,~R.~B.; Novaes,~F.~D.; Fazzio,~A.; da~Silva,~A. J.~R. Adsorption of
  Benzene-1,4-dithiol on the Au(111) Surface and Its Possible Role in Molecular
  Conductance. \emph{J. Am. Chem. Soc.} \textbf{2006}, \emph{128}, 8996--8997,
  PMID: 16834348\relax
\mciteBstWouldAddEndPuncttrue
\mciteSetBstMidEndSepPunct{\mcitedefaultmidpunct}
{\mcitedefaultendpunct}{\mcitedefaultseppunct}\relax
\EndOfBibitem
\bibitem[Cossaro et~al.({2008})Cossaro, Mazzarello, Rousseau, Casalis, Verdini,
  Kohlmeyer, Floreano, Scandolo, Morgante, Klein, and Scoles]{Cossaro:2008}
Cossaro,~A.; Mazzarello,~R.; Rousseau,~R.; Casalis,~L.; Verdini,~A.;
  Kohlmeyer,~A.; Floreano,~L.; Scandolo,~S.; Morgante,~A.; Klein,~M.~L. et~al.
  {X-ray Diffraction and Computation Yield the Structure of Alkanethiols on
  Gold(111)}. \emph{{Science}} \textbf{{2008}}, \emph{{321}}, {943--946}\relax
\mciteBstWouldAddEndPuncttrue
\mciteSetBstMidEndSepPunct{\mcitedefaultmidpunct}
{\mcitedefaultendpunct}{\mcitedefaultseppunct}\relax
\EndOfBibitem
\bibitem[Humphrey et~al.({1996})Humphrey, Dalke, and Schulten]{Humphrey:1996}
Humphrey,~W.; Dalke,~A.; Schulten,~K. {VMD: Visual Molecular Dynamics}.
  \emph{{J. Mol. Graph.}} \textbf{{1996}}, \emph{{14}}, {33--38}\relax
\mciteBstWouldAddEndPuncttrue
\mciteSetBstMidEndSepPunct{\mcitedefaultmidpunct}
{\mcitedefaultendpunct}{\mcitedefaultseppunct}\relax
\EndOfBibitem
\bibitem[Hostetler et~al.({1998})Hostetler, Wingate, Zhong, Harris, Vachet,
  Clark, Londono, Green, Stokes, Wignall, Glish, Porter, Evans, and
  Murray]{Hostetler:1998}
Hostetler,~M.; Wingate,~J.; Zhong,~C.; Harris,~J.; Vachet,~R.; Clark,~M.;
  Londono,~J.; Green,~S.; Stokes,~J.; Wignall,~G. et~al. {Alkanethiolate Gold
  Cluster Molecules with Core Diameters from 1.5 to 5.2 nm: Core and Monolayer
  Properties as a Function of Core Size}. \emph{{Langmuir}} \textbf{{1998}},
  \emph{{14}}, {17--30}\relax
\mciteBstWouldAddEndPuncttrue
\mciteSetBstMidEndSepPunct{\mcitedefaultmidpunct}
{\mcitedefaultendpunct}{\mcitedefaultseppunct}\relax
\EndOfBibitem
\bibitem[Horiguchi et~al.({2009})Horiguchi, Tsutsui, Kurokawa, and
  Sakai]{Horiguchi:2009}
Horiguchi,~K.; Tsutsui,~M.; Kurokawa,~S.; Sakai,~A. {Electron Transmission
  Characteristics of Au/1,4-Benzenedithiol/Au Junctions}. \emph{{Nanotech.}}
  \textbf{{2009}}, \emph{{20}}, {025204}\relax
\mciteBstWouldAddEndPuncttrue
\mciteSetBstMidEndSepPunct{\mcitedefaultmidpunct}
{\mcitedefaultendpunct}{\mcitedefaultseppunct}\relax
\EndOfBibitem
\bibitem[Haiss et~al.({2003})Haiss, van Zalinge, Higgins, Bethell, Hobenreich,
  Schiffrin, and Nichols]{Haiss:2003}
Haiss,~W.; van Zalinge,~H.; Higgins,~S.; Bethell,~D.; Hobenreich,~H.;
  Schiffrin,~D.; Nichols,~R. {Redox State Dependence of Single Molecule
  Conductivity}. \emph{{J. Am. Chem. Soc.}} \textbf{{2003}}, \emph{{125}},
  {15294--15295}\relax
\mciteBstWouldAddEndPuncttrue
\mciteSetBstMidEndSepPunct{\mcitedefaultmidpunct}
{\mcitedefaultendpunct}{\mcitedefaultseppunct}\relax
\EndOfBibitem
\bibitem[Cui et~al.({2001})Cui, Primak, Zarate, Tomfohr, Sankey, Moore, Moore,
  Gust, Harris, and Lindsay]{Cui:2001}
Cui,~X.; Primak,~A.; Zarate,~X.; Tomfohr,~J.; Sankey,~O.; Moore,~A.; Moore,~T.;
  Gust,~D.; Harris,~G.; Lindsay,~S. {Reproducible Measurement of
  Single-Molecule Conductivity}. \emph{{Science}} \textbf{{2001}},
  \emph{{294}}, {571--574}\relax
\mciteBstWouldAddEndPuncttrue
\mciteSetBstMidEndSepPunct{\mcitedefaultmidpunct}
{\mcitedefaultendpunct}{\mcitedefaultseppunct}\relax
\EndOfBibitem
\bibitem[Asar et~al.({2009})Asar, Mariscal, and Leiva]{Asar:2009}
Asar,~J. A.~O.; Mariscal,~M.~M.; Leiva,~E. P.~M. {Stochastic Model for
  Spontaneous Formation of Molecular Wires}. \emph{{Electrochim. Acta}}
  \textbf{{2009}}, \emph{{54}}, {2977--2982}\relax
\mciteBstWouldAddEndPuncttrue
\mciteSetBstMidEndSepPunct{\mcitedefaultmidpunct}
{\mcitedefaultendpunct}{\mcitedefaultseppunct}\relax
\EndOfBibitem
\bibitem[Kushmerick et~al.({2003})Kushmerick, Naciri, Yang, and
  Shashidhar]{Kushmerick:2003}
Kushmerick,~J.; Naciri,~J.; Yang,~J.; Shashidhar,~R. {Conductance Scaling of
  Molecular Wires in Parallel}. \emph{{Nano Lett.}} \textbf{{2003}},
  \emph{{3}}, {897--900}\relax
\mciteBstWouldAddEndPuncttrue
\mciteSetBstMidEndSepPunct{\mcitedefaultmidpunct}
{\mcitedefaultendpunct}{\mcitedefaultseppunct}\relax
\EndOfBibitem
\bibitem[Bratkovsky and Kornilovitch({2003})]{Bratkovsky:2003}
Bratkovsky,~A.; Kornilovitch,~P. {Effects of Gating and Contact Geometry on
  Current through Conjugated Molecules Covalently Bonded to Electrodes}.
  \emph{{Phys. Rev. B}} \textbf{{2003}}, \emph{{67}}, {115307}\relax
\mciteBstWouldAddEndPuncttrue
\mciteSetBstMidEndSepPunct{\mcitedefaultmidpunct}
{\mcitedefaultendpunct}{\mcitedefaultseppunct}\relax
\EndOfBibitem
\bibitem[Selzer et~al.({2004})Selzer, Cabassi, Mayer, and Allara]{Selzer:2004}
Selzer,~Y.; Cabassi,~M.; Mayer,~T.; Allara,~D. {Thermally Activated Conduction
  in Molecular Junctions}. \emph{{J. Am. Chem. Soc.}} \textbf{{2004}},
  \emph{{126}}, {4052--4053}\relax
\mciteBstWouldAddEndPuncttrue
\mciteSetBstMidEndSepPunct{\mcitedefaultmidpunct}
{\mcitedefaultendpunct}{\mcitedefaultseppunct}\relax
\EndOfBibitem
\bibitem[Chen et~al.({2006})Chen, Li, Hihath, Huang, and Tao]{Chen:2006}
Chen,~F.; Li,~X.; Hihath,~J.; Huang,~Z.; Tao,~N. {Effect of Anchoring Groups on
  Single-Molecule Conductance: Comparative Study of Thiol-, Amine-, and
  Carboxylic-Acid-Terminated Molecules}. \emph{{J. Am. Chem. Soc.}}
  \textbf{{2006}}, \emph{{128}}, {15874--15881}\relax
\mciteBstWouldAddEndPuncttrue
\mciteSetBstMidEndSepPunct{\mcitedefaultmidpunct}
{\mcitedefaultendpunct}{\mcitedefaultseppunct}\relax
\EndOfBibitem
\bibitem[Kruger et~al.({2002})Kruger, Fuchs, Rousseau, Marx, and
  Parrinello]{Kruger:2002}
Kruger,~D.; Fuchs,~H.; Rousseau,~R.; Marx,~D.; Parrinello,~M. {Pulling
  Monatomic Gold Wires with Single Molecules: An Ab Initio Simulation}.
  \emph{{Phys. Rev. Lett.}} \textbf{{2002}}, \emph{{89}}, {186402}\relax
\mciteBstWouldAddEndPuncttrue
\mciteSetBstMidEndSepPunct{\mcitedefaultmidpunct}
{\mcitedefaultendpunct}{\mcitedefaultseppunct}\relax
\EndOfBibitem
\bibitem[Kim et~al.(2011)Kim, Hellmuth, Bürkle, Pauly, and
  Scheer]{Kim-acs-nano:2011}
Kim,~Y.; Hellmuth,~T.~J.; Bürkle,~M.; Pauly,~F.; Scheer,~E. Characteristics
  of Amine-Ended and Thiol-Ended Alkane Single-Molecule Junctions Revealed by
  Inelastic Electron Tunneling Spectroscopy. \emph{ACS Nano} \textbf{2011},
  \emph{5}, 4104--4111\relax
\mciteBstWouldAddEndPuncttrue
\mciteSetBstMidEndSepPunct{\mcitedefaultmidpunct}
{\mcitedefaultendpunct}{\mcitedefaultseppunct}\relax
\EndOfBibitem
\bibitem[Iacovella et~al.({2011})Iacovella, French, and
  Cummings]{Iacovella_scidac:2011}
Iacovella,~C.; French,~W.; Cummings,~P. {Flexible Order Parameters for
  Quantifying the Rate-Dependent Energy Release Mechanism of Au Nanowires}.
  \emph{{Proc. SciDAC}} \textbf{{2011}},  {Denver, CO, July 10--14}\relax
\mciteBstWouldAddEndPuncttrue
\mciteSetBstMidEndSepPunct{\mcitedefaultmidpunct}
{\mcitedefaultendpunct}{\mcitedefaultseppunct}\relax
\EndOfBibitem
\bibitem[Plimpton({1995})]{Plimpton:1995}
Plimpton,~S. {Fast Parallel Algoritms for Short-Range Molecular-Dynamics}.
  \emph{{J. Comp. Phys.}} \textbf{{1995}}, \emph{{117}}, {1--19}\relax
\mciteBstWouldAddEndPuncttrue
\mciteSetBstMidEndSepPunct{\mcitedefaultmidpunct}
{\mcitedefaultendpunct}{\mcitedefaultseppunct}\relax
\EndOfBibitem
\bibitem[Cleri and Rosato({1993})]{Cleri:1993}
Cleri,~F.; Rosato,~V. {Tight-Binding Potentials for Transition-Metals and
  Alloys}. \emph{{Phys. Rev. B}} \textbf{{1993}}, \emph{{48}}, {22--33}\relax
\mciteBstWouldAddEndPuncttrue
\mciteSetBstMidEndSepPunct{\mcitedefaultmidpunct}
{\mcitedefaultendpunct}{\mcitedefaultseppunct}\relax
\EndOfBibitem
\bibitem[Pu et~al.({2007})Pu, Leng, Tsetseris, Park, Pantelides, and
  Cummings]{Pu:2007}
Pu,~Q.; Leng,~Y.; Tsetseris,~L.; Park,~H.~S.; Pantelides,~S.~T.;
  Cummings,~P.~T. {Molecular Dynamics Simulations of Stretched Gold Nanowires:
  The Relative Utility of Different Semiempirical Potentials}. \emph{{J. Chem.
  Phys.}} \textbf{{2007}}, \emph{{126}}, {144707}\relax
\mciteBstWouldAddEndPuncttrue
\mciteSetBstMidEndSepPunct{\mcitedefaultmidpunct}
{\mcitedefaultendpunct}{\mcitedefaultseppunct}\relax
\EndOfBibitem
\bibitem[Pu et~al.({2007})Pu, Leng, Zhao, and Cummings]{Pu-propane:2007}
Pu,~Q.; Leng,~Y.; Zhao,~X.; Cummings,~P.~T. {Molecular Simulations of
  Stretching Gold Nanowires in Solvents}. \emph{{Nanotech.}} \textbf{{2007}},
  \emph{{18}}, {424007}\relax
\mciteBstWouldAddEndPuncttrue
\mciteSetBstMidEndSepPunct{\mcitedefaultmidpunct}
{\mcitedefaultendpunct}{\mcitedefaultseppunct}\relax
\EndOfBibitem
\bibitem[Rappe et~al.({1992})Rappe, Casewit, Colwell, Goddard, and
  Skiff]{Rappe:1992}
Rappe,~A.; Casewit,~C.; Colwell,~K.; Goddard,~W.; Skiff,~W. {UFF, a Full
  Periodic-Table Force-Field For Molecular Mechanics and Molecular-Dynamics
  Simulations}. \emph{{J. Am. Chem. Soc.}} \textbf{{1992}}, \emph{{114}},
  {10024--10035}\relax
\mciteBstWouldAddEndPuncttrue
\mciteSetBstMidEndSepPunct{\mcitedefaultmidpunct}
{\mcitedefaultendpunct}{\mcitedefaultseppunct}\relax
\EndOfBibitem
\end{mcitethebibliography}

\pagebreak

\section{Graphical Table of Contents}

\begin{figure}[h!]
	\centering
	\includegraphics[width=6.0in]{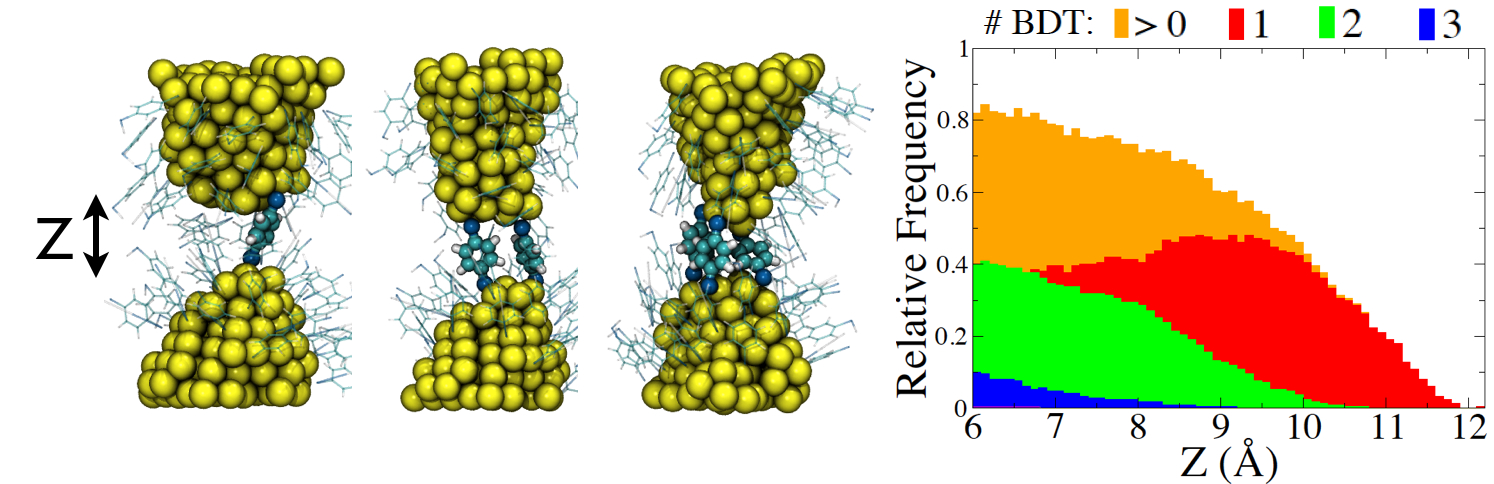}
	\label{fig:toc}
\end{figure}

\end{document}